\documentclass{jfm}
\usepackage{graphicx}
\usepackage{amssymb,amsmath,bm,accents}
\usepackage{natbib}
\usepackage{mathtools}

\DeclareMathSymbol{\widehatsym}{\mathord}{largesymbols}{"62}
\newcommand\lowerwidehatsym{%
  \text{\smash{\raisebox{-1.3ex}{%
    $\widehatsym$}}}}
\newcommand\hatt[1]{%
  \mathchoice
    {\accentset{\displaystyle\lowerwidehatsym}{#1}}
    {\accentset{\textstyle\lowerwidehatsym}{#1}}
    {\accentset{\scriptstyle\lowerwidehatsym}{#1}}
    {\accentset{\scriptscriptstyle\lowerwidehatsym}{#1}}
}

\renewcommand*\vec[1]{{\boldsymbol{#1}}}

\usepackage{etoolbox}
\makeatletter
\patchcmd{\l@section}
  {\hfil}
  {\leaders\hbox{\normalfont$\m@th\mkern \@dotsep mu\hbox{.}\mkern \@dotsep mu$}\hfill}
  {}{}
\makeatother

\makeatletter
\DeclareFontFamily{OMX}{MnSymbolE}{}
\DeclareSymbolFont{MnLargeSymbols}{OMX}{MnSymbolE}{m}{n}
\SetSymbolFont{MnLargeSymbols}{bold}{OMX}{MnSymbolE}{b}{n}
\DeclareFontShape{OMX}{MnSymbolE}{m}{n}{
    <-6>  MnSymbolE5
   <6-7>  MnSymbolE6
   <7-8>  MnSymbolE7
   <8-9>  MnSymbolE8
   <9-10> MnSymbolE9
  <10-12> MnSymbolE10
  <12->   MnSymbolE12
}{}
\DeclareFontShape{OMX}{MnSymbolE}{b}{n}{
    <-6>  MnSymbolE-Bold5
   <6-7>  MnSymbolE-Bold6
   <7-8>  MnSymbolE-Bold7
   <8-9>  MnSymbolE-Bold8
   <9-10> MnSymbolE-Bold9
  <10-12> MnSymbolE-Bold10
  <12->   MnSymbolE-Bold12
}{}

\let\llangle\@undefined
\let\rrangle\@undefined
\DeclareMathDelimiter{\llangle}{\mathopen}%
                     {MnLargeSymbols}{'164}{MnLargeSymbols}{'164}
\DeclareMathDelimiter{\rrangle}{\mathclose}%
                     {MnLargeSymbols}{'171}{MnLargeSymbols}{'171}
\makeatother

\DeclarePairedDelimiterX\braket[2]{\langle}{\rangle}{#1 \delimsize\vert #2}
\DeclarePairedDelimiterX\inner[2]{\langle}{\rangle}{#1,#2}
\DeclarePairedDelimiterX\minner[2]{\llangle}{\rrangle}{#1,#2}
\DeclarePairedDelimiter\abs{\lvert}{\rvert}
\DeclarePairedDelimiter\norm{\lVert}{\rVert}

\DeclarePairedDelimiter\avg{\langle}{\rangle}

\newcommand*\Ro{\mathrm{Ro}}

\renewcommand*\Re{\mathrm{Re}}

\usepackage{subcaption}

\usepackage[section]{placeins}
\usepackage{textcomp}
\usepackage{gensymb} 

\def\-{\raisebox{.75pt}{-}} 

\DeclareMathOperator{\sgn}{sgn}
\usepackage{float}
\usepackage{upgreek}
\usepackage[normalem]{ulem}

\newdimen\uulinesep
\makeatletter
\renewcommand*{\uuline}{%
  \bgroup
  \UL@setULdepth
  \markoverwith{%
    \lower\ULdepth\hbox{%
      \kern-.03em%
      \vtop{%
        \hrule width.2em%
        \kern\uulinesep
        \hrule
      }%
      \kern-.03em%
    }%
  }%
  \ULon
}
\makeatother

\DeclareMathSymbol{\upLambda}{\mathalpha}{operators}{3}
\usepackage{amsfonts,amsbsy,amsmath}
\usepackage{graphicx}
\usepackage{color}
\usepackage{natbib}
\usepackage{pgfplots}
\pgfplotsset{compat=1.5}

\usepackage{etoolbox}
\usepackage{upgreek}
\usepackage{lineno}

\newcommand*\patchAmsMathEnvironmentForLineno[1]{%
  \expandafter\let\csname old#1\expandafter\endcsname\csname #1\endcsname
  \expandafter\let\csname oldend#1\expandafter\endcsname\csname end#1\endcsname
  \renewenvironment{#1}%
     {\linenomath\csname old#1\endcsname}%
     {\csname oldend#1\endcsname\endlinenomath}}%
\newcommand*\patchBothAmsMathEnvironmentsForLineno[1]{%
  \patchAmsMathEnvironmentForLineno{#1}%
  \patchAmsMathEnvironmentForLineno{#1*}}%
\AtBeginDocument{%
\patchBothAmsMathEnvironmentsForLineno{equation}%
\patchBothAmsMathEnvironmentsForLineno{align}%
\patchBothAmsMathEnvironmentsForLineno{flalign}%
\patchBothAmsMathEnvironmentsForLineno{alignat}%
\patchBothAmsMathEnvironmentsForLineno{gather}%
\patchBothAmsMathEnvironmentsForLineno{multline}%
}

\DeclareMathAlphabet{\mathsfit}{T1}{\sfdefault}{\mddefault}{\sldefault}

\def\i{\mathrm{i}}

\def\d{\mathrm{d}}

\def\beq{\begin{equation}}
\def\eeq{\end{equation}}

\def\eps{\varepsilon}

\def\sgn{\mathrm{sgn} \,}

\def\M{M}

\def\P{\mathsf{P}}

\DeclareMathAlphabet{\mathsfbfit}{T1}{\sfdefault}{bx}{sl}

\def\L{{\mathsfbfit{{L}}}}
\def\NN{{\mathsfbfit{N}}}
\def\N{{\mathsfit{N}}}
\def\NNhat{{\widehat{\mathsfbfit{N}}}}
\def\Nhat{{\widehat{\mathsfit{N}}}}

\def\UUhat{{\widehat{\mathsfbfit{U}}}}
\def\Uhat{{\widehat{\mathsfit{U}}}}
\def\M{{\mathsfbfit{M}}}
\def\MM{{\mathsfit{M}}}
\def\Wwide{{\mathsfbfit{{W}}}}

\def\E{{\mathsfbfit{E}}}

\def\XX{{\mathsfbfit{X}}}
\def\YY{{\mathsfbfit{Y}}}

\makeatletter
\newcommand{\W}[1][0.75]{%
  \mathpalette\@W{#1}%
}
\newcommand*{\@W}[2]{%
  \scalebox{#2}[1]{$#1{\Wwide}\m@th$}%
}
\makeatother
\def\What{\widehat{\W}}

\newcommand{\bs}[1]{\boldsymbol{#1}}

\newcommand{\av}[1]{\langle #1 \rangle}

\def\e{\mathrm{e}}
\def\sigmatens{\bm{\upsigma}}

\title[Inertia-gravity-wave scattering]{Inertia-gravity-wave scattering by geostrophic turbulence}
\author[M. A. C. Savva, H. A. Kafiabad and J. Vanneste]{M. A. C. Savva, H. A. Kafiabad and J. Vanneste}
\affiliation{School of Mathematics and Maxwell Institute for Mathematical Sciences, \\
University of Edinburgh, Edinburgh, UK}

\begin{document}

\maketitle

\begin{abstract}
In rotating stratified flows including in the atmosphere and ocean, inertia-gravity waves (IGWs) often coexist with a geostrophically balanced turbulent flow. Advection and refraction by this flow lead to wave scattering, redistributing  IGW energy in the position--wavenumber phase space. We give a detailed description of this process by deriving a kinetic equation governing the evolution of the IGW phase-space energy density. The derivation relies on the smallness of the Rossby number characterising the geostrophic flow, which is treated as a random field with known statistics, and makes no assumption of spatial scale separation.

The kinetic equation describes energy transfers that are restricted to IGWs with the same frequency, as a result of the timescale separation between waves and flow. We formulate the kinetic equation on the constant-frequency surface -- a double cone in wavenumber space -- using polar spherical coordinates, and we examine the form of the two scattering cross sections involved,  which quantify energy transfers between IGWs with, respectively, the same and opposite directions of vertical propagation. The kinetic equation captures both the horizontal isotropisation and the cascade of energy across scales that result from scattering. We focus our attention on the latter to assess the predictions of the kinetic equation against direct simulations of the three-dimensional Boussinesq equations, finding good agreement.

\end{abstract}

\section{Introduction}

This paper develops a statistical theory for the impact of a turbulent flow on the propagation of inertia-gravity waves (IGWs).
It is motivated, broadly, by the importance of IGWs for the circulation of both the atmosphere and ocean and, specifically, by two strands of research. The first centres around the decomposition of atmospheric and oceanic energy spectra into IGWs and quasigeostrophic flow. Analyses of aircraft \citep{cali-et-al14,cali-et-al16}, ship-track \citep{buhl-et-al14, roch-et-al16, buhler_kuang_tabak_2017} and simulation \citep{qiu-et-al-2018-seasonality,torres2018partitioning} data indicate that IGWs exist at larger scales and at greater energies than previously thought. They suggest that IGWs dominate over the quasigeostrophic flow across the broad range of horizontal scales (from 500 km down in the atmosphere, from 100 km in the ocean) characterised by a shallow kinetic energy spectrum, traditionally interpreted as a ${-5/3}$ power law in the atmosphere \citep{nast-gage} and a $-2$ power law in the ocean \citep{cali-ferr}. While this dominance remains controversial \citep{li-linb,asselin2018,kafi-bart}, the importance of IGWs at these scales is widely accepted.

These results raise basic questions about the processes that control the distribution of IGW energy in this range. One key process is the advection and refraction of the IGWs by the typically highly-energetic quasigeostrophic flow. The present paper gives a full description of the scattering of IGW energy that results from this advection and refraction.  We achieve this by applying powerful techniques of the theory of waves in random media \citep{ryzhik,powe-v,bal-et-al} to obtain a kinetic equation governing the evolution of  the IGW energy density, denoted by $a(\vec{x},\vec{k},t)$, in the position--wavenumber $(\vec{x},\vec{k})$ phase space. The main assumption is that the quasigeostrophic flow can be represented as a space- and time-dependent, homogeneous and stationary random field with known statistics.

We obtained partial results in this direction in a previous paper \citep{kafiabad_savva_vanneste_2019}
which focusses on the WKBJ regime, where the IGW scales are asymptotically smaller than the flow scale \citep[see][for earlier work]{muller76,muller77,watson1984-internalWaves,mull-et-al}. In that case, the Doppler shift of the IGW frequency resulting from advection by the flow is the sole mechanism of scattering and it acts as a diffusion in $\vec{k}$-space. A remarkable prediction in this diffusive regime is that the energy spectrum of forced IGWs equilibrates to a $k^{-2}$ power law for scales smaller than the forcing scale, similar to the spectra observed in the atmosphere and ocean. The present paper extends the WKBJ results by relaxing the assumption of separation between wave and flow scales, treating the distinguished limit when both are similar.
The scattering is then described by an integral operator which reduces to a diffusion only in the WKBJ limit. Earlier work by \citet{danioux} and
\citet{savva_vanneste_2018} derived and studied the  scattering operator relevant to, respectively, near-inertial waves and IGWs under the restriction of a barotropic ($z$-independent) quasigeostrophic flow. The results we obtain for fully three-dimensional flows are markedly different because vertical shear leads to a cascade of IGWs to small scales that is absent for barotropic flows.

The second strand of research motivating this paper is concerned with fundamental aspects of turbulence in rotating stratified flows  and specifically with their analysis in terms of triadic interactions. The interactions between two IGW modes and a geostrophic (or vortical) mode have been examined by \citet{warn1986statistical}, \citet{lelong}, \citet{bartello} and more recently by \cite{ward} and \citet{wagner_ferrando_young_2017}. They are often termed `catalytic' interactions because they leave the geostrophic mode unaffected, a property that stems from potential-vorticity conservation. Our results provide a  statistical description of precisely those catalytic interactions, with detailed predictions for the IGW spectrum that emerges in both initial-value and forced scenarios. A key aspect is that we confine our predictions to the statistics of the IGWs, regarding the statistics of the geostrophic modes as given. This is natural since the latter are determined by fully nonlinear (quasigeostrophic) turbulence and not amenable to the type of asymptotic treatment used for the IGWs. This is also partly justified by the catalytic nature of the IGW--geostrophic mode interactions which implies that the feedback of the IGWs on the geostrophic flow is weak. This feedback is what is captured by the theory of wave--mean flow interactions, in particular the generalised Lagrangian mean theory of \citet{andrews1978exact} \citep[see also][]{buhler_2014,Wagner2015AvailableFlow,gilbert2018geometric,kafiabad_vanneste_young2020}.
We note that the kinetic equation that we derive is closely related to the kinetic equations of wave (or weak) turbulence theory \citep[e.g.][]{naza}: the integral terms with quadratic nonlinearity obtained in wave turbulence for triadic interactions simplify to linear integrals when the amplitudes of one type of modes -- here the geostrophic modes -- remains fixed as we assume. Wave-turbulence theory provides a useful description of the interactions between IGWs \citep[e.g.][and references therein]{lvov-et-al} but ignores the effect of the quasigeostrophic flow or assumes it consists of a superposition of Rossby waves \citep{eden-chouksey-olbers-2019} and so its predictions are complementary to those of this paper.

As emphasised above, the statistical theory that we develop  makes no assumption of spatial scale separation between IGWs and the quasigeostrophic flow. As a result, its central object, namely the phase-space energy density $a(\vec{x},\vec{k},t)$, cannot be defined using a straighforward ray-tracing, WKBJ treatment. Instead, we follow \citet{ryzhik} and use the Wigner transform to both define $a(\vec{x},\vec{k},t)$ and obtain an equation governing its evolution \citep[see][for other applications of the Wigner transform to IGWs]{onuki2020quasi}. The kinetic equation governing the evolution of $a(\vec{x},\vec{k},t)$ is derived in \S\ref{sec:kinetic} and Appendix \ref{app:kineticderivation} and takes the form
\begin{equation}
  \partial_ta(\vec{x},\vec{k},t)+\nabla_{\vec{k}}\omega(\vec{k}) \cdot\nabla_{\vec{x}}a(\vec{x},\vec{k},t)  =\int_{\mathbb{R}^3}\sigma(\vec{k},\vec{k}')a(\vec{x},\vec{k}',t) \, \d\vec{k}'-\Sigma(\vec{k})a(\vec{x}, \vec{k},t). \label{eq:kineq_full}
\end{equation}
Here $\omega(\vec{k})$ is the IGW dispersion relation, $\sigma(\vec{k},\vec{k}')$ is the scattering cross section, which fully encodes the impact of the geostrophic flow on IGWs and is given explicitly in \eqref{eq:IGW-full-cross-section}, and $\Sigma(\vec{k}) = \int_{\mathbb{R}^3} \sigma(\vec{k},\vec{k}') \, \d \vec{k}'$.

A key property of $\sigma(\vec{k},\vec{k}')$ is that it is proportional to $\delta \left(\omega(\vec{k}) - \omega(\vec{k}') \right)$. This stems from the slow time dependence of the quasigeostrophic flow and implies that energy transfers between IGWs are restricted to waves with the same frequency. These waves have wavevectors lying on a double cone whose two halves, termed nappes, make angles $\theta=\tan^{-1} \left (\omega^2-f^2)/(N^2-\omega^2)\right)^{1/2}$ and $\pi-\theta$ with the $k_3$-axis ($f$ and $N$ are the inertial and buoyancy frequencies), corresponding to upward- and downward-propagating waves. In \S\ref{sec:cone} we reformulate the kinetic equation \eqref{eq:kineq_full} in spherical coordinates $(k,\theta,\varphi)$ well suited to the geometry of the constant-frequency cone since $\theta$ can be regarded as a fixed parameter. We then separate the energy density $a(\vec{x},\vec{k},t)$  into upward- and downward-propagating components and obtain a pair of coupled kinetic equations governing their evolution.
We examine the properties of these equations in some detail in \S\ref{sec:cone} and show, in particular, how they predict the isotropisation of the IGW field and the equipartition of energy between upward- and downward-propagating IGWs in the long-time limit.

In \S\ref{sec:IGWsimulations} we compare the predictions of the kinetic equations with results of high-resolution numerical simulations of the three-dimensional Boussinesq equations for an initial-value problem. We focus on homogeneous and horizontally isotropic configurations, when $a(\vec{x},\vec{k},t)$  is independent of $\vec{x}$ and of the azimuthal angle $\varphi$, and find very good agreement for different IGW frequencies and geostrophic-flow strengths. We briefly discuss the forced problem and confirm that the stationary spectrum that emerges has the $k^{-2}$ power tail behaviour
expected from the WKBJ, diffusive approximation of \citet{kafiabad_savva_vanneste_2019}.

\section{Kinetic equation} \label{sec:kinetic}

\subsection{Fluid-dynamical model}

We model the propagation of IGWs through a turbulent quasigeostrophic eddy field using the inviscid non-hydrostatic Boussinesq equations linearised about a background flow. The background flow depends slowly on time, is in geostrophic and hydrostatic balance and accordingly determined by a streamfunction $\psi$. We take $\psi$ to be a random field with  homogeneous and stationary statistics. The background flow velocity and buoyancy fields are  given by $\vec{U}=(U,V,0)=(-\partial_y\psi,\partial_x\psi,0)$ and $B=f\partial_z\psi$, and the linearised  Boussinesq equations read
\begin{subequations} \label{boussinesq}
\begin{align}
       {\partial_t\vec{u}}+\nabla\vec{U}\cdot\vec{u}+\vec{U}\cdot\nabla\vec{u}+f\hat{\vec{z}}\times\vec{u} &=-\nabla p+b\hat{\vec{z}},  \\
    {\partial_t b}+\vec{u}\cdot\nabla B+\vec{U}\cdot\nabla b+N^2w &=0,\\
    \nabla\cdot\vec{u}&=0,
\end{align}
\end{subequations}
where $\vec{u}=(u,v,w)$ denotes the wave velocity, $\nabla=(\partial_x,\partial_y,\partial_z)$ is the full gradient operator, $\hat{\vec{z}}$ is the vertical unit vector, $p$ is the wave pressure normalised by a constant reference density, $b$ the wave buoyancy, $f$ the Coriolis parameter, and $N$ the buoyancy frequency which is assumed to be constant with $N > f$.

Among the five equations in \eqref{boussinesq}, only three are prognostic since two of the five dependent variables $(u,v,w,b,p)$, e.g.\ $w$ and $p$, can be diagnosed from the remaining three. We make this explicit by reformulating \eqref{boussinesq} using three suitable dependent variables chosen as the linearised ageostrophic vertical vorticity, horizontal divergence and linearised potential vorticity
\begin{equation}
\gamma = f\zeta-\nabla_\mathrm{h}^2 \nabla^{-2} \left(\partial_zb-f\zeta\right), \quad
       \delta=\partial_x u+\partial_y v  \quad \textrm{and} \quad
       q=f\partial_zb+N^2\zeta,
\end{equation}
with $\nabla_\mathrm{h}=(\partial_x,\partial_y,0)$ and $\zeta = \partial_x v - \partial_y u$, following \cite{Vanneste2013BalanceFlows}. Since the potential vorticity $q$ describes the dynamics of the balanced flow which, in our formulation, is captured by the background flow, we set $q=0$. This reduces the dynamics to the two equations
\begin{subequations} \label{eq:new_eqns_motion}
\begin{align}
  {\partial_t\gamma}+{\Omega}^2\delta &=\mathcal{N}_{\gamma},\\
 {\partial_t \delta}-\gamma &=\mathcal{N}_\delta, \end{align}
\end{subequations}
where $\Omega$ is the pseudodifferential operator
\begin{equation}
\Omega(\nabla)=[(N^2\nabla_\mathrm{h}^2+f^2\partial_{zz})\nabla^{-2}]^{1/2}
\label{eq:IGW_dispersion_pseudo_operator}
\end{equation}
and $\mathcal{N}_\gamma$ and $\mathcal{N}_\delta$ groups the terms depending on the background flow. When these are ignored, the solutions to \eqref{eq:new_eqns_motion} can be written as superposition of plane IGWs, with wavevectors $\vec{k}=(\vec{k}_\mathrm{h},k_3)$ and frequencies
\begin{equation}
    \omega(\vec{k})= \pm \sqrt{N^2{k}_\mathrm{h}^2+f^2k_3^2}/{|\vec{k}|}\label{eq:IGW_dispersion_relation},
\end{equation}
with $k_\mathrm{h} = |\vec{k}_\mathrm{h}|$.

We now make some scaling assumptions. Our main assumption is that the Rossby number characterising the background flow is small:
\begin{equation}
\Ro = U_* K_* / f \ll 1,
\end{equation}
where $U_*$ and $K_*^{-1}$ are characteristic velocity and horizontal length scales of the flow. This assumption is consistent with the assumed geostrophic balance. It ensures that advection and refraction of the IGWs by the background flow are weak compared to wave dispersion. We also assume that the background flow evolves on a time scale $(\Ro f)^{-1}$ as is the case for quasigeostrophic dynamics. Crucially we make no assumption of separation of spatial scales and consider instead the distinguished regime where flow and IGWs have horizontal scales that are similar, $k_\mathrm{h}/K_*=O(1)$.

To make the scaling assumptions explicit while retaining the practical dimensional form of the  equations of motion, we introduce a bookkeeping parameter $\varepsilon \ll 1$ indicating the dependence of the various terms on powers of $\Ro$. A convenient choice takes $\varepsilon = \Ro^2$ since it turns out that the temporal and spatial variations of the IGW amplitudes then scale as $(\varepsilon \omega)^{-1}$ and $(\eps K_*)^{-1}$. With this choice we rewrite \eqref{eq:new_eqns_motion} in the compact form
\begin{equation}
  \partial_t\vec{\phi}+\L(\nabla)\vec{\phi }+\eps^{1/2}\NN(\vec{x},\nabla,\eps^{1/2}t)\vec{\phi}=0,
  \label{eq:IGW-general-eqns}
\end{equation}
where
\begin{equation}
\vec{\phi} = \left(\begin{array}{c}\gamma \\ \delta\end{array}\right)
\label{IGW_phi}
\end{equation}
groups the dynamical variables, and
\begin{equation}
  \L(\nabla)=\left(\begin{array}{cc}
0 & \Omega^2\\
-1 & 0
\end{array}\right). \label{eq:IGW_operator_L}
\end{equation}
The (matrix) linear operator $\NN$ collects the background-flow terms. It depends on $\vec{x}$ and $\eps^{1/2} t$ through the streamfunction $\psi$ and is given explicitly as \eqref{eq:N} in Appendix \ref{app:kineticderivation}. We next exploit the smallness of $\varepsilon$ to derive a kinetic equation governing the slow energy exchanges among IGWs resulting from interactions with the background flow.

\subsection{Derivation of the kinetic equation}

We start by rescaling space and time according to $(\vec{x},t) \mapsto (\vec{x}/\eps,t/\eps)$ so that $\vec{x}$ and $t$ capture the slow variations of the IGW amplitudes; the IGW phases then vary with $\vec{x}/\eps$ and $t/\eps$, and the background flow with $\vec{x}/\eps$ and $t/\sqrt{\eps}$. The rescaling transforms \eqref{eq:IGW-general-eqns} into
\begin{equation}
    \varepsilon\partial_t\vec{\phi}+\L(\varepsilon\nabla) \vec{\phi}+\eps^{1/2}\NN({\vec{x}}/{\varepsilon},\varepsilon\nabla,{t}/{\eps^{1/2}})\vec{\phi}=0.
    \label{eq:scaled-equations}
\end{equation}

A key ingredient for the systematic derivation of the kinetic equation is the definition of a phase-space energy (or action) density $a(\vec{x},\vec{k},t)$ that does not rest on the WKBJ approximation. The separation between spatial and wavenumber information required for a phase-space description of the waves is achieved by means of the (scaled) Wigner transform of $\vec{\phi}$ defined as the $2 \times 2$ matrix
\begin{equation}
    \W (\vec{x},\vec{k},t)= \frac{1}{(2\pi)^3} \int_{\mathbb{R}^3}\e^{\i \vec{k}\cdot\vec{y}}\vec{\phi}(\vec{x}-\varepsilon{\vec{y}}/{2},t)\vec{\phi}^\mathrm{T}(\vec{x}+\varepsilon{\vec{y}}/{2},t) \, {\d\vec{y}},
    \label{eq:scaled-wigner}
\end{equation}
where $\mathrm{T}$ denotes the transpose. In Appendix \ref{app:kineticderivation} we derive an evolution equation for $\W$ which we simplify using multiscale asymptotics. The derivation starts with the expansion
\begin{equation}
    \W=\W^{(0)}(\vec{x},\vec{k},t)+\eps^{1/2}\W^{(1)}(\vec{x},\boldsymbol\xi,\vec{k},t,\tau)+\varepsilon \W^{(2)}(\vec{x},\boldsymbol\xi,\vec{k},t,\tau)+{O}(\varepsilon^{3/2}),\label{wigner_multiscale}
\end{equation}
where $\vec{\xi} = \vec{x}/\eps$ and $\tau=t/\eps^{1/2}$ are treated as independent of $\vec{x}$ and $t$. The leading-order equation obtained is
\begin{equation}
\L(\i\vec{k})\W^{(0)} + \mathrm{c.c.} = 0,
\label{eq:LOW}
\end{equation}
where, from \eqref{eq:IGW_operator_L},
\begin{equation}
    \L(\i\vec{k})=\left(\begin{array}{cc}
        0 & \omega^2(\vec{k}) \\
        -1 & 0
    \end{array}\right),\end{equation}
with $\omega(\vec{k})$ the IGW frequency given by \eqref{eq:IGW_dispersion_relation}.
This matrix has eigenvalues $\pm \i \omega(\vec{k})$ and eigenvectors $\vec{e}_\pm$ solving
\begin{equation}
\L(\i \vec{k})\vec{e}_\pm(\vec{k})=\pm\i\omega(\vec{k})\vec{e}_\pm(\vec{k}),
\label{eq:evectors}
\end{equation}
where we choose $\omega(\vec{k})>0$ by convention.
The eigenvectors encode the polarisation relations of IGWs. They can be written as
\begin{equation}
    \vec{e}_\pm(\vec{k})=\frac{|\vec{k}_\mathrm{h}||k_3|}{\sqrt{2}\,|\vec{k}|}\left(\begin{array}{c}
        \pm\i\omega(\vec{k})   \\
         -1
    \end{array}\right)\label{eq:igw_eigenvectors}
\end{equation}
and are orthonormal with respect to a weighted inner-product, specifically
\begin{equation}
    \inner{\vec{e}_{i}(\vec{k})}{\vec{e}_{j}(\vec{k})}_\M=\vec{e}_{i}^*(\vec{k})\M\vec{e}_{j}(\vec{k})=\delta_{ij},\label{igw_inner}
\end{equation}
where  the symmetric matrix $\M$ is defined by
\begin{equation}
    \M(\vec{k})=\frac{|\vec{k}|^2}{\omega^2|\vec{k}_\mathrm{h}|^2|k_3|^2}\left(\begin{array}{cc}
        1 & 0 \\
        0 & \omega^2(\vec{k})
    \end{array}\right).\label{igw_matrix_M}
\end{equation}
Eq.\ \eqref{eq:LOW} is solved in terms of the eigenvectors $\vec{e}_\pm(\vec{k})$: defining
the matrices
\begin{equation}
    \E_j(\vec{k})=\vec{e}_j(\vec{k})\vec{e}^*_j(\vec{k}),
\end{equation}
the solution reads
\begin{equation}
    \W^{(0)}(\vec{x},\vec{k},t)=\sum_{j=\pm}a_j(\vec{x},\vec{k},t)\E_j(\vec{k}) \label{W02}
\end{equation}
for amplitudes  $a_j(\vec{x},\vec{k},t)$ to be determined. Because, by definition \eqref{eq:scaled-wigner}, $\W$ is Hermitian, these amplitudes are real. The reality of $\vec{\phi}$ further implies that $\W(\vec{x},-\vec{k},t)= \W^\mathrm{T}(\vec{x},\vec{k},t)$ and hence $a_+(\vec{x},-\vec{k},t)=a_-(\vec{x},\vec{k},t)$. We can therefore focus on a single amplitude, $a(\vec{x},\vec{k},t) = a_+(\vec{x},\vec{k},t)$, say. This is the desired phase-space energy density. This interpretation is justified by the fact that its integral over $\vec{k}$ approximates the energy density:
\begin{align}
\mathcal{E}(\vec{x},t) &= \tfrac{1}{2} \left( |\vec{u}|^2 + {b^2}/{N^2} \right) = \tfrac{1}{2} \left( \M^{1/2}(\eps \i \nabla) \vec{\phi} \right)^\mathrm{T} \left( \M^{1/2}(\eps \i \nabla) \vec{\phi} \right) \nonumber \\
&= \tfrac{1}{2} \int_{\mathbb{R}^3} \mathrm{tr} \left( \M(\vec{k}) \W^{(0)}(\vec{x},\vec{k},t) \right) \d \vec{k} + O(\eps) = \int_{\mathbb{R}^3} a(\vec{x},\vec{k},t) \, \d \vec{k} + O(\eps).
\end{align}

An evolution equation for $a(\vec{x},\vec{k},t)$ is derived by considering higher-order terms in the expansion \eqref{wigner_multiscale} and imposing a solvability condition as detailed in Appendix \ref{app:kineticderivation}. The result is the kinetic equation \eqref{eq:kineq_full} with the differential scattering cross section
\begin{align}
  \sigma(\vec{k},\vec{k}')=&\frac{\pi |k_3|^2|k_3'|^2}{2\omega^4 |\vec{k}|^2 |\vec{k}'|^2 |\vec{k}_\mathrm{h}|^2 |\vec{k}_\mathrm{h}'|^2}\Bigg[  |\vec{k}_\mathrm{h}'\times\vec{k}_\mathrm{h}|^2\big[(N^2+\omega^2)\frac{|\vec{k}_\mathrm{h}|^2|\vec{k}_\mathrm{h}'|^2}{|k_3| |k_3'|}\sgn(k_3k_3')\nonumber \\
  &+(f^2+\omega^2)(2\vec{k}_\mathrm{h}\cdot\vec{k}_\mathrm{h}'-|\vec{k}_\mathrm{h}||\vec{k}_\mathrm{h}'|\sgn(k_3k_3'))\big]^2+f^2\omega^2\Big(2|\vec{k}_\mathrm{h}'\times\vec{k}_\mathrm{h}|^2 \label{eq:IGW-full-cross-section} \\
  &+\big[|\vec{k}_\mathrm{h}'-\vec{k}_\mathrm{h}|^2-(k_3'-k_3)^2\frac{|\vec{k}_\mathrm{h}'||\vec{k}_\mathrm{h}| }{|k_3'||k_3|}  \big]\vec{k}_\mathrm{h}'\cdot\vec{k}_\mathrm{h}  \Big)^2 \Bigg]\frac{\hatt{E}_\mathrm{K}(\vec{k}'-\vec{k})}{|\vec{k}_\mathrm{h}'-\vec{k}_\mathrm{h}|^2}\delta\left(\omega(\vec{k}')-\omega(\vec{k})\right), \nonumber
\end{align}
where $\hatt{E}_\mathrm{K}(\vec{k})$ is the kinetic energy spectrum of the background geostrophic flow, and $\Sigma(\vec{k})$ is the total scattering cross section
\begin{equation}
\Sigma(\vec{k})  = \int_{\mathbb{R}^3} \sigma(\vec{k},\vec{k}') \,  \d \vec{k}'.
\end{equation}

The cross section \eqref{eq:IGW-full-cross-section} is the principal object of interest and first main result of this paper. It fully quantifies the impact that scattering by a quasigeostrophic turbulent flow has on the statistics of IGWs.
Before analysing this impact in detail, we make four remarks.
\begin{enumerate}
\item The obvious symmetry $\sigma(\vec{k},\vec{k}') = \sigma(\vec{k}',\vec{k})$ ensures that the scattering is energy conserving: the energy density
\begin{equation}
  \mathcal{E}_0(\vec{x},t)=\int_{\mathbb{R}^3} a(\vec{x},\vec{k},t) \, \d\vec{k}
\end{equation}
satisfies the conservation law
\begin{equation}
  \partial_t\mathcal{E}_0+\nabla_{\vec{x}}\cdot\vec{\mathcal{F}}_0=0,
  \label{eq:energycons}
\end{equation}
with the flux
\begin{equation}
  \vec{\mathcal{F}}_0(\vec{x},t)=\int_{\mathbb{R}^3}\nabla_{\vec{k}}\omega(\vec{k}) \, a(\vec{x},\vec{k},t) \, \d\vec{k}.
\end{equation}
Conservation of the volume-integrated energy follows. We emphasise that this conservation is not trivial. The  Boussinesq equations linearised about a background flow \eqref{boussinesq} do not conserve the perturbation energy, even when the flow is time independent. The conservation law \eqref{eq:energycons} arises from the phase averaging implicit in the definition of $a(\vec{x},\vec{k},t)$ and, in this sense, should be interpreted as an action conservation law. Energy and action are equivalent to the level of accuracy of our approximation because the Doppler shift is a factor $\eps^{1/2}$ smaller than the intrinsic frequency of the IGWs.

\item The factor $\delta\left(\omega(\vec{k}')-\omega(\vec{k})\right)$ indicates that the energy exchanges caused by scattering are restricted to a constant-frequency surface in $\vec{k}$-space, that is, the cone $k_3/k_\mathrm{h} = \mathrm{const}$. This is because the evolution of the background flow is slow enough that the flow is treated as time independent. Scattering then results from resonant triadic interactions in which one mode -- the catalyst vortical mode -- has zero frequency, and the other two modes -- the IGWs -- have equal and opposite frequencies. The geometry of the constant-frequency cone is crucial to the nature of the scattering. In the next section we account for it explicitly by rewriting the kinetic equation on the constant-frequency cone itself, using spherical polar coordinates.

\item We can connect \eqref{eq:IGW-full-cross-section} to earlier results on the scattering of IGWs by a barotropic (i.e, $z$-independent) quasigeostrophic flow \citep{savva_vanneste_2018}. The assumption of a barotropic flow amounts to taking $
\hatt{E}_\mathrm{K}(\vec{k}) = \hatt{E}_\mathrm{K,B}(\vec{k}_\mathrm{h}) \delta(k_3)$, which implies that
$k_3=k_3'$ in \eqref{eq:IGW-full-cross-section}, hence $|\vec{k}_\mathrm{h}|=|\vec{k}_\mathrm{h}'|$ and $|\vec{k}|=|\vec{k}'|$ in view of the resonance condition $\omega(\vec{k})=\omega(\vec{k}')$. If we further make the hydrostatic approximation $|\vec{k}|\approx |k_3|$ \citep{olbers} we obtain
\begin{align}
  \sigma(\vec{k},\vec{k}')&=\frac{2\pi }{\omega^4  |\vec{k_\mathrm{h}}|^4 }\Bigg( |\vec{k}_\mathrm{h}'\times\vec{k}_\mathrm{h}|^2\big((\omega^2+f^2)\vec{k}_\mathrm{h}\cdot\vec{k}_\mathrm{h}'-f^2|\vec{k}_\mathrm{h}|^2\big)^2\label{eq:IGW-IT-sigma-approx}\\
  &+f^2\omega^2\Big(|\vec{k}_\mathrm{h}'\times\vec{k}_\mathrm{h}|^2 +\vec{k}_\mathrm{h}\cdot\vec{k}_\mathrm{h}'\big(|\vec{k}_\mathrm{h}|^2-\vec{k}_\mathrm{h}\cdot\vec{k}_\mathrm{h}'\big) \Big)^2 \Bigg)\frac{\hatt{E}_{\mathrm{K,B}}(\vec{k}'_\mathrm{h}-\vec{k}_\mathrm{h})}{|\vec{k}_\mathrm{h}'-\vec{k}_\mathrm{h}|^2}\delta(\omega(\vec{k}')-\omega(\vec{k})), \nonumber
\end{align}
which is identical to the cross section derived for the rotating shallow-water system in \citet{savva_vanneste_2018}. It is further shown in that paper that the cross section reduces to that derived for near-inertial waves by \citet{danioux} when $\omega \to f$.

\item The WKBJ limit of the kinetic equation is obtained by assuming that the energy of the quasigeostrophic flow is concentrated at scales large compared with the wave scales; formally, $\hatt{E}_\mathrm{K}(\vec{k}) = g(\alpha^{-1} \vec{k})$ for $\alpha \ll 1$ and some function $g$ that decreases rapidly for large argument. In this limit, it can be shown that the scattering terms in \eqref{eq:kineq_full} reduce to the (wavenumber) diffusion derived by \citet{kafiabad_savva_vanneste_2019} taking the WKBJ approximation as a starting point \cite[see][for details]{savva2020}.
This makes it clear that the results of the present paper extend those of \citet{kafiabad_savva_vanneste_2019} to capture a broad range of wave scales, from scales larger than the quasigeostrophic-flow scales down to arbitrarily small scales.

\end{enumerate}

We next rewrite the kinetic equation \eqref{eq:kineq_full} in a form tailored to the geometry of the constant-frequency cones on which the energy exchanges are restricted and discuss its properties.

\section{Scattering on the constant-frequency cone} \label{sec:cone}

\subsection{Kinetic equation in spherical coordinates}

 We use spherical polar coordinates for the wavenumbers, writing

\begin{equation}
    \vec{k}=k\left(\begin{array}{c}
        \sin\theta\cos\varphi   \\
         \sin\theta\sin\varphi\\
         \cos\theta
    \end{array}\right)
    \quad \textrm{and} \quad
     \vec{k}'={k}'\left(\begin{array}{c}
        \sin\theta'\cos(\varphi+\varphi')   \\
         \sin\theta'\sin(\varphi+\varphi')\\
         \cos\theta'
    \end{array}\right).\label{coordinates}
\end{equation}
Note that we use $\varphi'$ for the difference between the azimuthal angles of wavevectors $\vec{k}'$ and $\vec{k}$ rather then the azimuthal angle of $\vec{k}'$ itself.
In these coordinates the dispersion relation \eqref{eq:IGW_dispersion_relation} reads
\begin{equation}
 \omega(\vec{k}) = \omega(\theta)=\sqrt{N^2\sin^2\theta+f^2\cos^2\theta}.
\end{equation}
The constant-frequency constraint $\omega(\theta')=\omega(\theta )$ implies that $\theta'=\theta$ or $\theta'=\pi-\theta$,
where
\begin{equation}
0 \le  \theta=\sin^{-1}\sqrt{\frac{\omega(\vec{k})^2-f^2}{N^2-f^2}} \le \pi/2
\end{equation}
is a constant. We interpret this as follows: the constant-frequency cone has two nappes, one corresponding to upward-propagating waves and the other to downward-propagating waves, and a wave of a certain type, upward-propagating say, exchanges energy with both upward- and downward propagating waves. We separate the two types of exchanges by writing
the delta function in \eqref{eq:IGW-full-cross-section} in the new coordinates \eqref{coordinates} as
\begin{align}
    \delta(\omega(\vec{k}')-\omega(\vec{k}))&=\frac{2\,\omega(\theta')}{|\sin(2\theta')|(N^2-f^2)}\left(\delta(\theta'-\theta)+\delta(\theta'-(\pi-\theta))\right)
    \label{eq:2deltas}
\end{align}
and defining the pair of cross sections $\sigma_\pm$ by
\begin{equation}
 \sum_{j=\pm}\sigma_j({k},{k}',\varphi, \varphi',\theta) =   \int_{0}^{\pi}\sigma\sin\theta'\d\theta', \label{sigmaprime}
\end{equation}
each associated with the contribution of a single $\delta$-function. In this way,  $\sigma_+$ quantifies the rate of scattering between waves on the same nappe of the constant-frequency cone, while $\sigma_-$ quantifies the rate of scattering between the two nappes and thus the wave reflection induced by interactions with the flow. Introducing \eqref{eq:2deltas} into \eqref{eq:IGW-full-cross-section} and carrying out the integration in $\theta'$ gives
\begin{multline}
    \sigma_\pm(k,k',\varphi,\varphi',\theta)=\frac{\pi{k}^2{k}'^2}{16\omega^3}\frac{|\sin(2\theta)|^3}{\sin\theta(N^2-f^2)}\left\{ 4f^2\omega^2\left[\cos\varphi'(\cos\varphi'\mp 1)-\sin^2\varphi'\right]^2  \right. \\
    \left. +\sin^2\varphi'\left[(\omega^2+f^2)(2\cos\varphi'\mp 1)\pm(N^2+\omega^2){\tan^2\theta} \right]^2 \right\}\frac{\hatt{E}_\mathrm{K}(\vec{k}'-\vec{k})}{({k}^2+{k}'^2-2{k}'{k}\cos\varphi')},\label{sigma_pm}
\end{multline}
where it is understood that $\vec{k}'$ in the argument of the spectrum $\hatt{E}_\mathrm{K}(\vec{k}'-\vec{k})$ is restricted to represent the set of wavevectors on the same nappe of the constant-frequency cone as $\vec{k}$ for $\sigma_+$ and on the opposite nappe for $\sigma_-$.

\begin{figure}
  \begin{subfigure}{1.05\linewidth}\vspace{.5cm}\hspace{-2em}
      \begin{subfigure}[t]{.49\linewidth}
        \centering
        \includegraphics[width=1\linewidth]{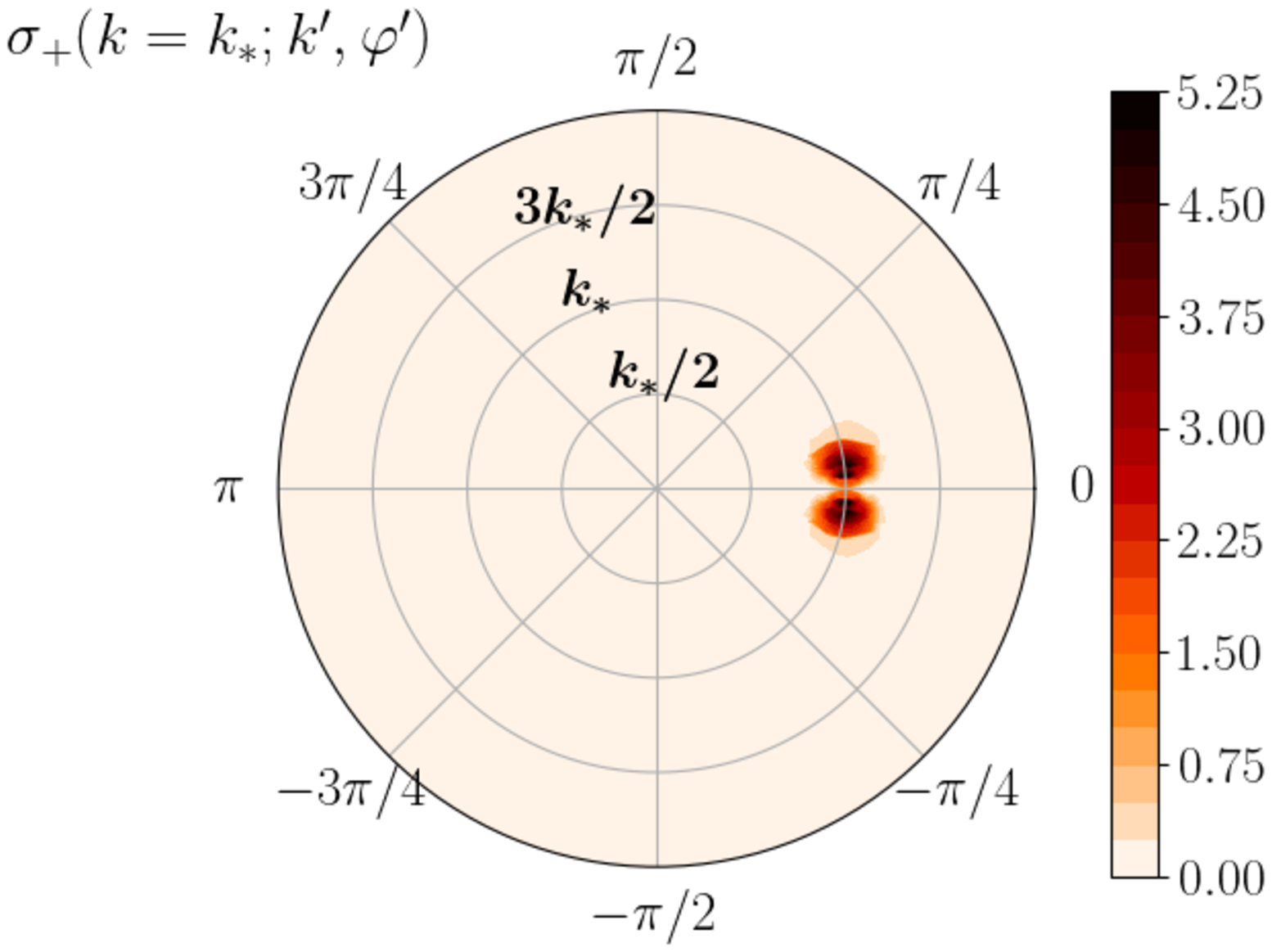}
      \end{subfigure}\hspace{.5em}
      \begin{subfigure}[t]{.49\linewidth}
        \centering
      \includegraphics[width=1\linewidth]{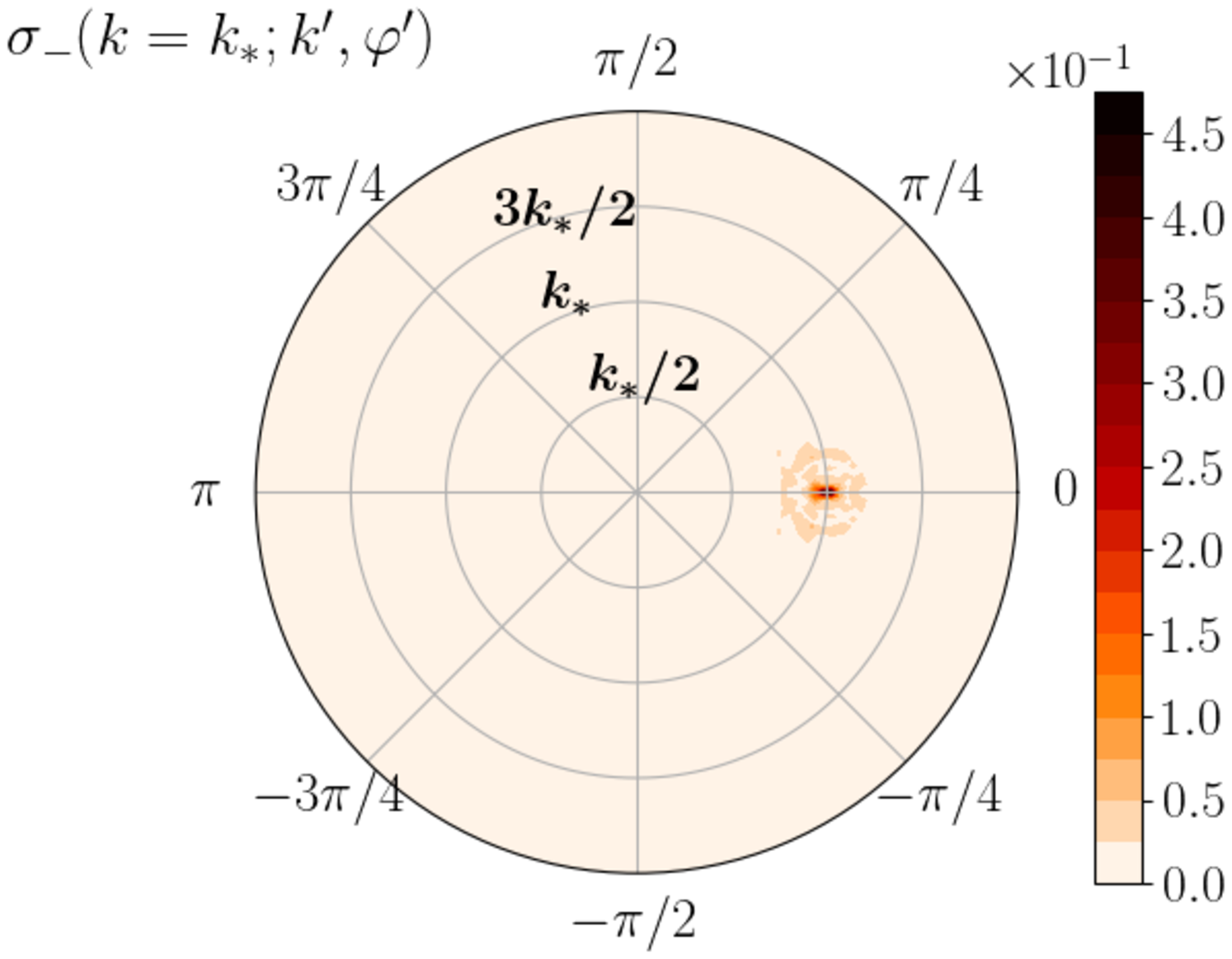}
    \end{subfigure}
\end{subfigure}

\hspace{-2em}
    \begin{subfigure}{1.05\linewidth}
      \begin{subfigure}[t]{.49\linewidth}
        \centering
        \includegraphics[width=1\linewidth]{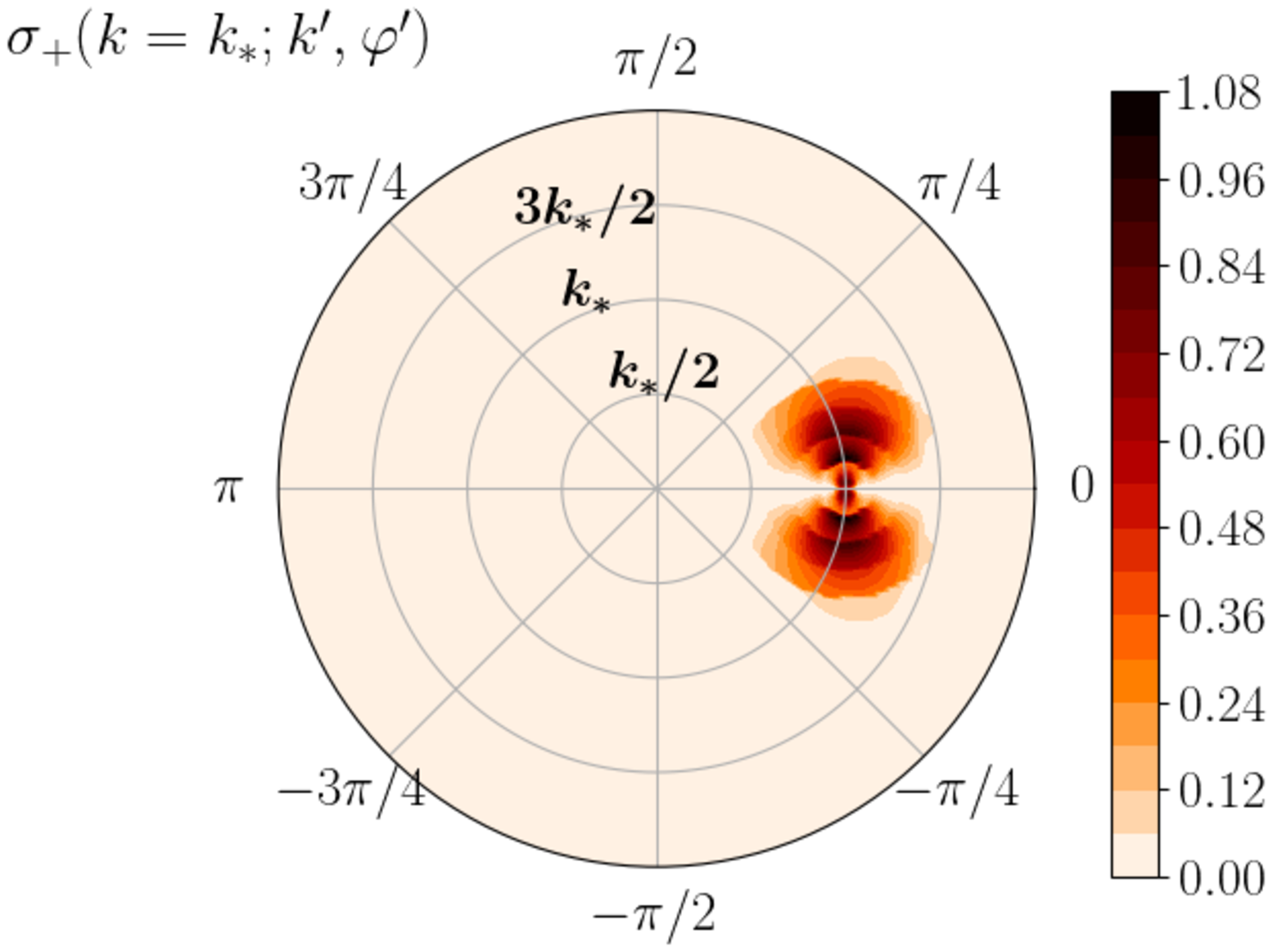}
      \end{subfigure}\hfill
      \begin{subfigure}[t]{.49\linewidth}
        \centering
        \includegraphics[width=1\linewidth]{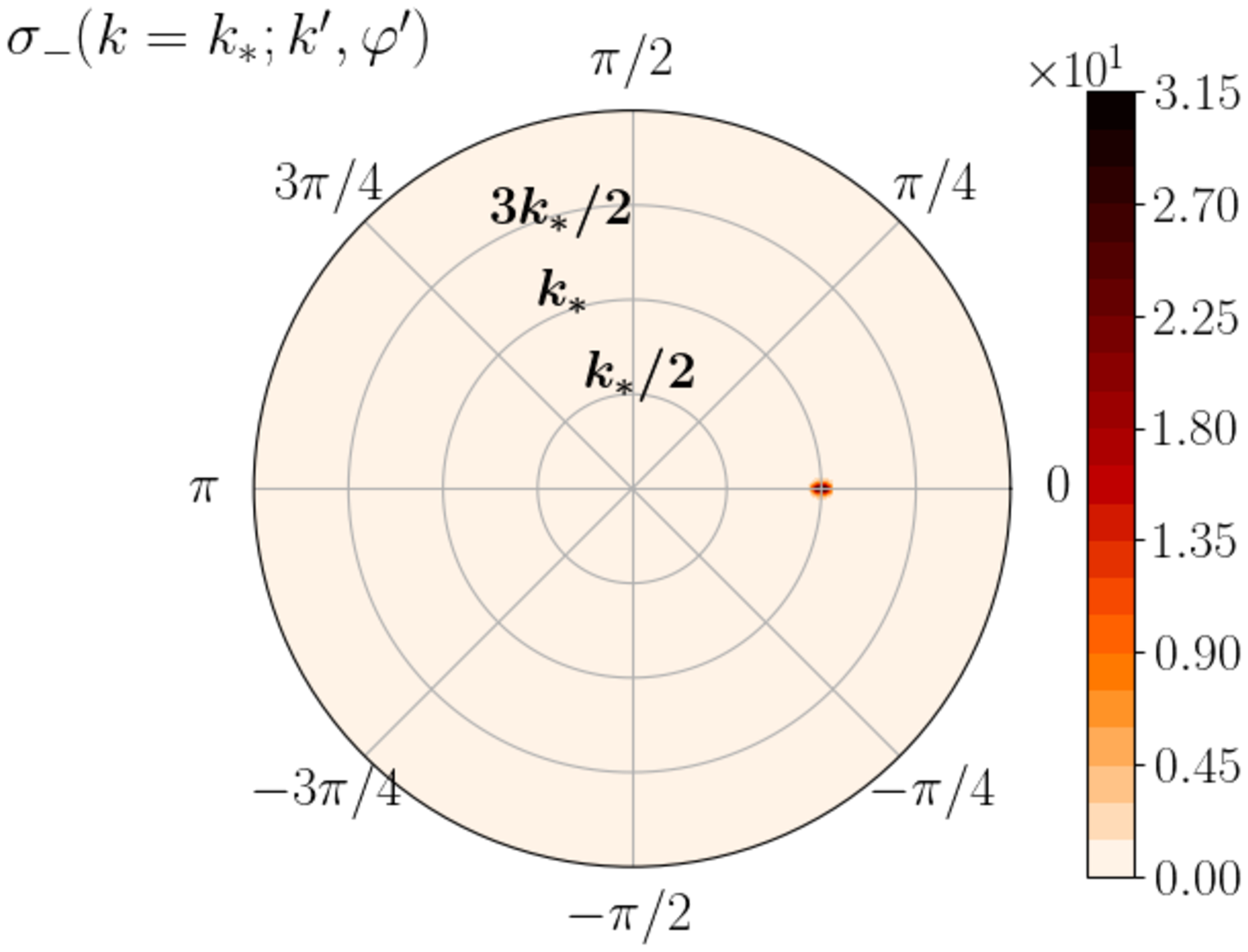}
      \end{subfigure}
    \end{subfigure}

  \centering\caption{Scattering cross sections $\sigma_\pm(k=k_*,k',\varphi')$ for $\omega=3f$, $\Ro=0.099$, $N/f=32$, and the quasigeostrophic-flow energy spectrum described in  \S\ref{sec:numerics}.
 The ratio of the IGW wavenumber to the geostrophic-flow peak wavenumber is $k_*/K_*\simeq 4$ (WKBJ regime, top panels) and $k_*/K_*\simeq 1$ (bottom panels).}\label{fig:IGWsigma}
\end{figure}

We emphasise that the cross sections $\sigma_\pm$ depend on the azimuthal angle $\varphi$ solely through the background-flow spectrum $\hatt{E}_\mathrm{K}$. This dependence disappears for horizontally isotropic flows and the cross sections are then functions of three variables only: $\sigma_\pm=\sigma_\pm(k,k',\varphi')$. We use this to illustrate the form of $\sigma_\pm$ for a fixed $k=k_*$ in Figure \ref{fig:IGWsigma}. The energy spectrum $\hatt{E}_\mathrm{K}$ used is obtained by azimuthally averaging the spectrum obtained in a geostrophic turbulence simulation described in \S\ref{sec:IGWsimulations}. The figure indicates that $\sigma_+$ is localised around $(k'=k_*, \varphi'=0)$. This implies spectrally local energy transfers and stems from the concentration of the background-flow energy at large scales. The localisation is increasingly marked as the ratio $k_*/K_*$ of the IGW wavenumber to the geostrophic-flow peak wavenumber increases. This culminates in the WKBJ regime $k_*/K_* \gg 1$, when scattering is well described by a fully-local diffusion in \citet{kafiabad_savva_vanneste_2019}. The broader support of $\sigma_+$ in $\varphi'$ compared to $k'$ suggests that scattering leads to a rapid wave energy spreading in the azimuthal direction, that is, a rapid isotropisation in the horizontal, followed by a slower radial spreading associated with a cascade towards small scales. Numerical simulations (not shown) confirm this general tendency.

\begin{figure}\centering
  \includegraphics[width=.5\linewidth]{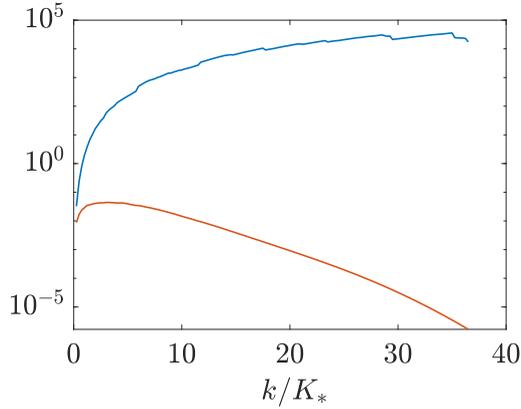}
  \caption{Total cross sections for same-nappe  and across-nappe  transfers $\Sigma_+(k)$ (blue line) and $\Sigma_-(k)$ (red line) defined in \eqref{eq:Sigpm} for $\omega=2f$, $N/f=32$, $\Ro=0.099$  and the quasigeostrophic-flow energy spectrum described in  \S\ref{sec:numerics}.    \label{fig:Sig_pm}
  }
\end{figure}

The corresponding plots of $\sigma_-$ in Figure \ref{fig:IGWsigma} indicate that the transfers between nappes of the constant-frequency cone are weak, especially for large $k_*/K_*$. Although the maximum pointwise  value of $\sigma_-$ can exceed that of $\sigma_+$ for  $k_*/K_*$ of order one, integrated values are more meaningful. We therefore show
\begin{equation}
\Sigma_\pm({k},\theta)=\int_0^\infty\int_{-\pi}^{\pi}\sigma_\pm(k,k',\varphi',\theta){k}'^2 \, \d\varphi'\d{k}',\label{eq:Sigpm}
\end{equation}
in Figure \ref{fig:Sig_pm} to confirm the dominance of $\sigma_+$ over $\sigma_-$ and hence of energy transfers on the same nappe of the cone over energy transfers between nappes. The values of $\sigma_-$ and $\Sigma_-$ decrease as $k_*/K_*$ increases, and in the WKBJ limit the transfers between nappes are completely negligible; in other words, short IGWs do not get reflected.

With the spherical polar coordinates, it is convenient to introduce the energy density
\begin{equation}
b(\vec{x},k,\varphi,\theta,t)=\sin\theta \, k^2 \, a(\vec{x},k,\varphi,\theta,t)
\end{equation}
such that $b(\vec{x},k,\varphi,\theta,t) \, \d k \d \varphi$ is the energy in $[k,k+\d k] \times [\varphi,\varphi + \d \varphi]$ and to partition it according to the nappe that it occupies (equivalently the direction of vertical propagation) by defining
\begin{equation}
  \vec{b}(\vec{x},{k},\varphi,t)  = \begin{pmatrix} b_+(\vec{x},{k},\varphi,t) \\ b_-(\vec{x},{k},\varphi,t) \end{pmatrix} =  \begin{pmatrix} b(\vec{x},{k},\varphi,\theta,t) \\ b(\vec{x},{k},\varphi,\pi-\theta,t) \end{pmatrix}
\quad \textrm{with} \ \ \theta \in [0,\pi/2].
\label{eq:vecb}
\end{equation}
We omit the parametric dependence on $\theta$ from now. With the definition \eqref{eq:vecb},  the kinetic equation \eqref{eq:kineq_full} becomes
\begin{subequations} \label{eq:kineq1}
\begin{align}
    {{\partial_t} \vec{b}}(\vec{x},\vec{k},t) &+ \nabla_{\vec{k}}\omega(\vec{k}) \cdot\nabla_{\vec{x}}\vec{b}(\vec{x},\vec{k},t) \\ &=k^2\iint\sigmatens({k},{k}',\varphi,\varphi')\, \vec{b}(\vec{x},{k}',\varphi-\varphi',t)\d{k}'\d\varphi'
    -\Sigma({k},\varphi)\,\vec{b}(\vec{x},{k},\varphi,t), \nonumber
\end{align}
where the matrix-valued cross section
\begin{equation}
    \sigmatens=\left(\begin{array}{cc}
        \sigma_+ & \sigma_- \\
        \sigma_- & \sigma_+
    \end{array}\right)\label{matrix_sigma}
\end{equation}
\end{subequations}
has components defined in \eqref{sigma_pm} and $\Sigma = \Sigma_+ + \Sigma_-$ follows from \eqref{eq:Sigpm}. Eq.\ \eqref{eq:kineq1}, consisting of a pair of coupled kinetic equations in the two-dimensional $(k,\varphi)$-space, provides the most useful description of the scattering of IGWs by geostrophic turbulence. It simplifies further for horizontally isotropic flows since the explicit dependence on $\varphi$ disappears and Fourier series can be employed. We discuss properties of the scattering inferred from  \eqref{eq:kineq1} in the next section.

\subsection{Properties of the scattering}\label{sec:IGW-kineq-properties}

The sum $b_++b_-$ of the two components of $\vec{b}$ is the total energy density and is conserved:
\begin{equation}
\mathcal{E}_0(\vec{x},t) = \int_0^\infty \int_{-\pi}^\pi \left( b_+(\vec{x},k,\varphi,t) +  b_-(\vec{x},k,\varphi,t) \right) \, \d k \d \varphi
\end{equation}
satisfies the conservation law \eqref{eq:energycons}. The difference $\Delta b = b_+-b_-$, on the other hand, can be shown to satisfy
\begin{align}
    \partial_t \int_0^\infty \int_{-\pi}^\pi \Delta b(\vec{x},k,\varphi,t) \, \d k\d \varphi\    =-2\int_0^\infty \int_{-\pi}^\pi \Sigma_-(k,\varphi)\Delta b(\vec{x},k,\varphi,t) \d  k\d\varphi,\label{eq:equilibration}
\end{align}
using the evenness of $\sigma_\pm$ in $\varphi'$ and the reversibility symmetry $\sigma'_\pm(k,k',\varphi,\varphi')=\sigma'_\pm(k',k,\varphi+\varphi',-\varphi')$. Since $\Sigma_- > 0$, this shows that $\iint \Delta b \, \d k \d \varphi$ decays with time at a rate controlled by the cross section $\Sigma_-$, so that the scattering leads to an equipartition between upward- and downward propagating IGWs. Note that twice the maximum of $\Sigma_-$, $2 \norm{\Sigma_-}_\infty$, provides a lower bound on the rate at which this equipartition occurs \cite[see][]{savva2020}.

In common with other kinetic equations, \eqref{eq:kineq_full} or \eqref{eq:kineq1} satisfy an H-theorem \citep{villani2008} showing that the entropy
\begin{equation}
   - \int_{\mathbb{R}^3} a\ln a \, \d\vec{k}\d\vec{x}
\end{equation}
increases. This implies that IGW energy spreads on the constant-energy cone in an irreversible manner. Because the cone is not compact, there is no possibility of reaching a steady state, so the scattering leads to a continued scale cascade, mostly towards small scales as a result of the cone geometry, that is only arrested by dissipation. This is in sharp contrast with the situation in the absence of vertical shear where the constant-frequency sets are circles (intersections of the cones with the surfaces $k_3 = \mathrm{const}$) and a steady state is reached, corresponding to an isotropic distribution of IGW energy when the flow is horizontally isotropic \citep{savva_vanneste_2018}.

We now focus on the case of an isotropic background flow, when the cross sections \eqref{sigma_pm} are independent of the azimuthal variable $\varphi$.  Expanding both sides of \eqref{eq:kineq1} in Fourier series gives
\begin{equation}
   \partial_t  \hatt{\vec{b}}_n(\vec{x},k,t) +\nabla_{\vec{k}}\omega(\vec{k}) \cdot\nabla_{\vec{x}}\hatt{\vec{b}}_n(\vec{x},k,t) ={2\pi}k^2\int_0^\infty{\hatt{\sigmatens}}\phantom{}_{n}(k,k')\,\hatt{\vec{b}}_n(\vec{x},{k}',t) \, \d{k}'    -\Sigma(k)\,\hatt{\vec{b}}_n(\vec{x},{k},t),\label{isotropic_wigner_3d}
\end{equation}
where the hats denote the Fourier coefficients defined as
\begin{equation}
    \hatt{\vec{b}}_n(\vec{x},k,t)=\frac{1}{2\pi}\int_{-\pi}^\pi\e^{\i n\varphi} \, \vec{b}(\vec{x},k,\varphi,t)\,\d\varphi.
\end{equation}
We can show from \eqref{isotropic_wigner_3d} that, for $n \not=0$, $\int \hatt{\vec{b}}_n \, \d k \to 0$ as $t \to \infty$.  This is seen by integrating \eqref{isotropic_wigner_3d} with respect to $k$ and summing the $\pm$ components of $\hatt{\vec{b}}_n$ to obtain
\begin{align}
   \partial_t & \int_0^\infty \left( \hatt{b}_{n+}(\vec{x},k,t)+\hatt{b}_{n-}(\vec{x},k,t) \right) \, \d k \nonumber \\
   &=-\int_0^\infty\big(\Sigma(k)-\Lambda_n(k)  \big) \left( \hatt{b}_{n+}(\vec{x},k,t)+\hatt{b}_{n-}(\vec{x},k,t) \right) \d k,\label{integral_isotropisation}
\end{align}
where
\begin{equation}
    \Lambda_n(k)=2\pi \int_0^\infty (\hatt{\sigma}_{n+}(k,k')+\hatt{\sigma}_{n-}(k,k')) k'^2 \, \d k'.\label{modal_scat_operator}
\end{equation}
It follows from \eqref{modal_scat_operator} and and the properties of  Fourier coefficients that
\begin{equation}
    \Lambda_0(k)=\Sigma(k) \;\;\; \text{and}\;\;\; |\Lambda_{n}(k)|<\Lambda_0(k) \ \ \textrm{for} \ \ n \ge 1.
\end{equation}
Thus the scattering term on the right-hand side of \eqref{integral_isotropisation} vanishes for $n=0$ and is negative for $n\geq 1$, so that the amplitudes $\hatt{b}_{n\pm}$ decay for all but the isotropic ($n=0$) mode. Hence the IGW wavefield  becomes horizontally isotropic in the long-time limit irrespective of the initial conditions.

In the remainder of the paper, we test the predictions of the kinetic equation \eqref{isotropic_wigner_3d} against direct numerical simulations of the Boussinesq equations. We focus on an initial condition that is approximately spatially homogeneous and horizontally isotropic so that the transport term $\nabla_{\vec{k}}\omega\cdot\nabla_{\vec{x}}\hatt{\vec{b}}_n$ can be neglected and only the mode $n=0$ needs to be considered.

\section{Kinetic equation vs Boussinesq simulations}\label{sec:IGWsimulations}

\subsection{Setup and numerical methods} \label{sec:numerics}

We carry out a set of Boussinseq simulations similar to those in \citet{kafiabad_savva_vanneste_2019}, using
a code adapted from that in \citet{wait-bart06b} based on a de-aliased pseudospectral method and a third-order Adams--Bashforth scheme with timestep $0.015/f$. The triply-periodic domain, $(2 \pi)^3$ in the scaled coordinates $(x,y,z'=Nz/f)$, is discretised uniformly with $768^3$ grid points and a hyperdissipation of the form $-\nu  (\partial_x^8+\partial_y^8 + \partial_{z'}^8)$, with $\nu = 2\times10^{-17}$ is added to the momentum and density equations.  We take $N/f=32$, a representative value of mid-depth ocean stratification. The initial condition is the superposition of a turbulent flow, obtained by running a quasigeostrophic model to a statistically stationary state, and IGWs. The initial spectrum of the flow peaks at $K_{\mathrm{h*}} \simeq 4$ and has an inertial subrange scaling approximately as $K_\mathrm{h}^{-3}$ and $K_\mathrm{3}^{-3}$.  This spectrum evolves slowly over the IGW-diffusion timescale, and its time-average defines $\hatt{E}_\mathrm{K}$ which is used to calculate the cross sections $\hatt{\sigmatens}_n({k},{k}')$ in \eqref{isotropic_wigner_3d}. IGWs are initialised along a ring in wavenumber space, with random phases and identical magnitudes, so that the corresponding spectrum is horizontally isotropic, that is, independent of $\varphi$. Since this remains (approximately) the case throughout the simulation, we concentrate on the evolution of the spectrum $\hatt{\vec{b}}_0(k,t)$ of the isotropic, $n=0$ mode.

Simulations are performed for two Rossby numbers $\Ro= K_{\mathrm{h}*} \av{|\bs{U}|^2}^{1/2}/f =0.049,\, 0.099$ (or $ \av{\zeta^2}^{1/2}/f=0.1,\, 0.2$ for the alternative Rossby numbers based on the vertical vorticity $\zeta$), which we refer to as `low' and `high' Rossby numbers, and the two IGW frequencies $\omega=2f, \, 3f$.
 We carry our experiments with  two different IGW horizontal wavenumbers: (i) $k_\mathrm{h*}=16\simeq 4 K_\mathrm{h*}$, as used in \citet{kafiabad_savva_vanneste_2019}, which is large enough to be in the WKBJ regime where the scattering integral in \eqref{isotropic_wigner_3d} reduces to a diffusion; and (ii) $k_\mathrm{h*}=4\simeq  K_\mathrm{h*}$ which requires the full kinetic equation.
The IGW energy spectrum  is computed at each step following the normal-mode decomposition of \citet{bartello}. We retain data for $(k_\mathrm{h},f k_3 /N)\in[0,254]\times[-255,255]$, and deduce the two components of
\begin{equation}
\hatt{\vec{b}}_0(k,t) = \frac{1}{2\pi} \int_{-\pi}^\pi \vec{b}(k,\varphi,t) \, \d \varphi
\end{equation}
 on a one-dimensional uniform grid with $k\in[-254/\sin\theta,254/\sin\theta]$ by projection onto the constant-frequency cone. Conventionally, we take negative values of $k$ for the lower nappe of the cone (i.e.\  $\pi/2 < \theta<\pi$) so $\hatt{b}_0(k,t)=\hatt{b}_{0+}(k,t)$ for $k \ge 0$ and $\hatt{b}_0(k,t)=\hatt{b}_{0-}(-k,t)$ for $k \le 0$.
In what follows, we omit the hat and subscript $0$ from $\hatt{b}_0(k,t)$.

We solve the kinetic equation \eqref{isotropic_wigner_3d} for the horizontally isotropic mode $n=0$  on an evenly-spaced grid
interpolated to provide twice the resolution of data from the Boussinesq simulations for a given frequency.
We interpolate the geostrophic kinetic-energy spectrum $\hatt{E}_\mathrm{K}$ to double the resolution in each dimension for computing the cross sections $\sigma_\pm$. We employ an FFT
to compute the $\varphi$-averaged cross section $\hat{\sigmatens}_0$.
Eq.\ \eqref{isotropic_wigner_3d} is integrated in time using an Euler scheme with timesteps chosen so that $\Delta t =0.5\,(\max_{{k}}\Sigma({k}))^{-1}$. The integrals in $k'$ are discretised  as Riemann sums,
which respects the energy conservation property of the kinetic equation.
Absorbing layers are used to prevent cascaded energy from building up.  For comparison, the diffusion equation of \citet{kafiabad_savva_vanneste_2019} is solved on the upper nappe on the grid ${k\in[0,254/\sin\theta]}$ with the same resolution as the kinetic equation, using first order central-difference differences for the $k$-derivatives, and a stiff ODE solver for time-stepping.

\subsection{Initial-value problem} \label{sec:IVP}

\begin{figure}
\begin{center}
\begin{tabular}{cc}
\includegraphics[width=.45\linewidth]{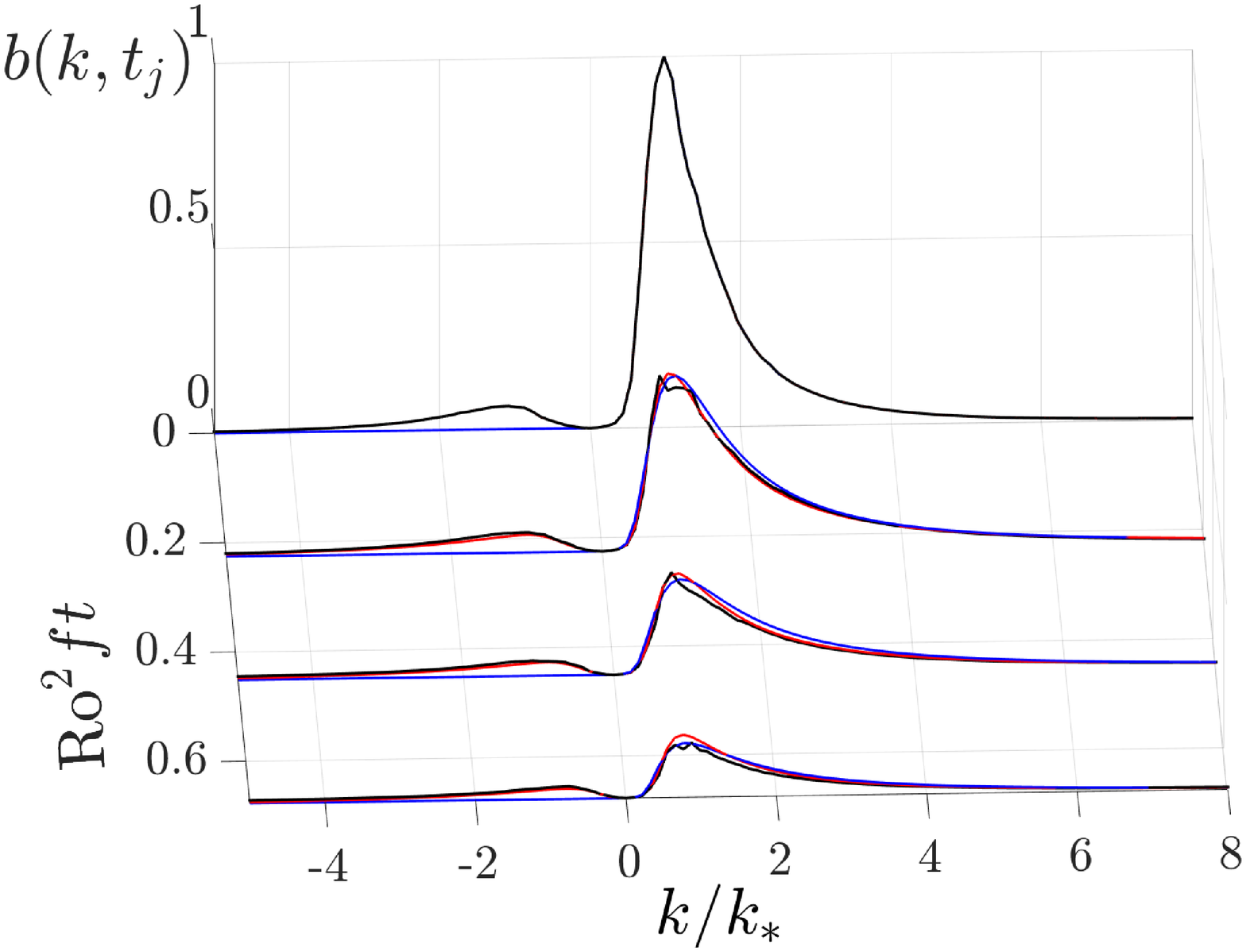} &
\includegraphics[width=.45\linewidth]{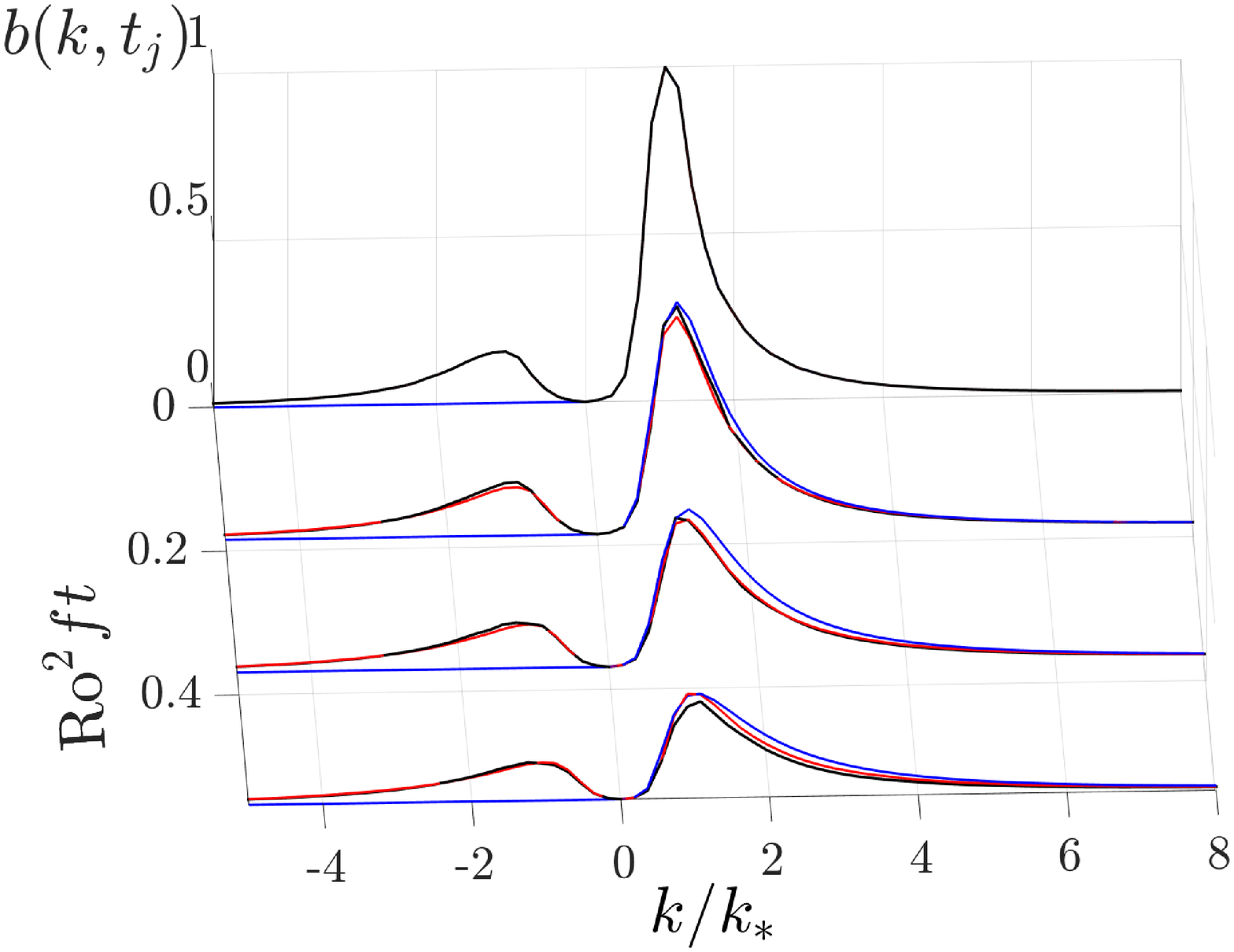} \\
{(a) $\omega=2f$, $\Ro=0.049$, $k_{\mathrm{h}*}/K_{\mathrm{h}*}\simeq 4$} & {(b) $\omega=3f$, $\Ro=0.099$, $k_{\mathrm{h}*}/K_{\mathrm{h}*}\simeq 4$} \bigskip \\
\includegraphics[width=.45\linewidth]{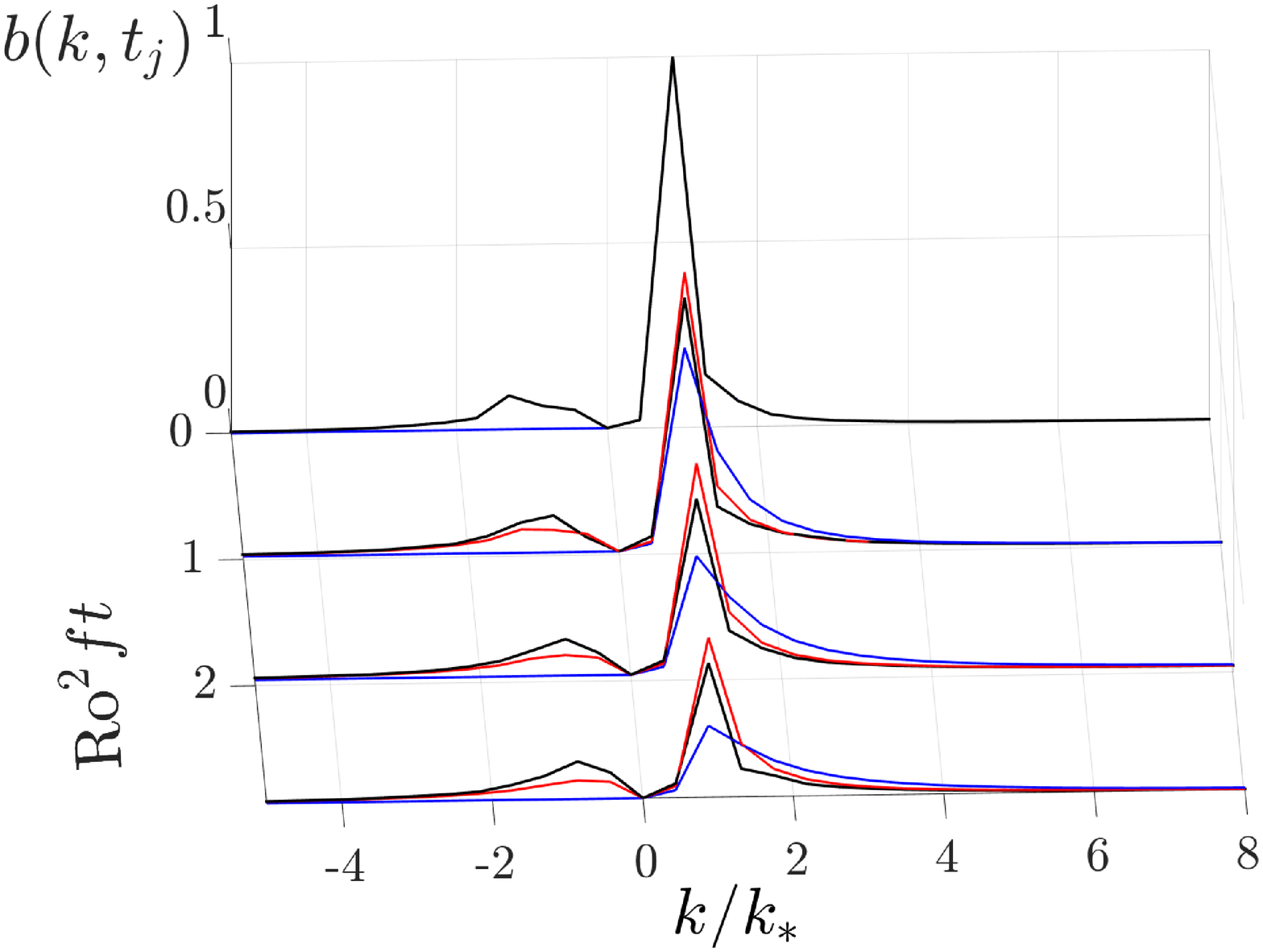} &
\includegraphics[width=.45\linewidth]{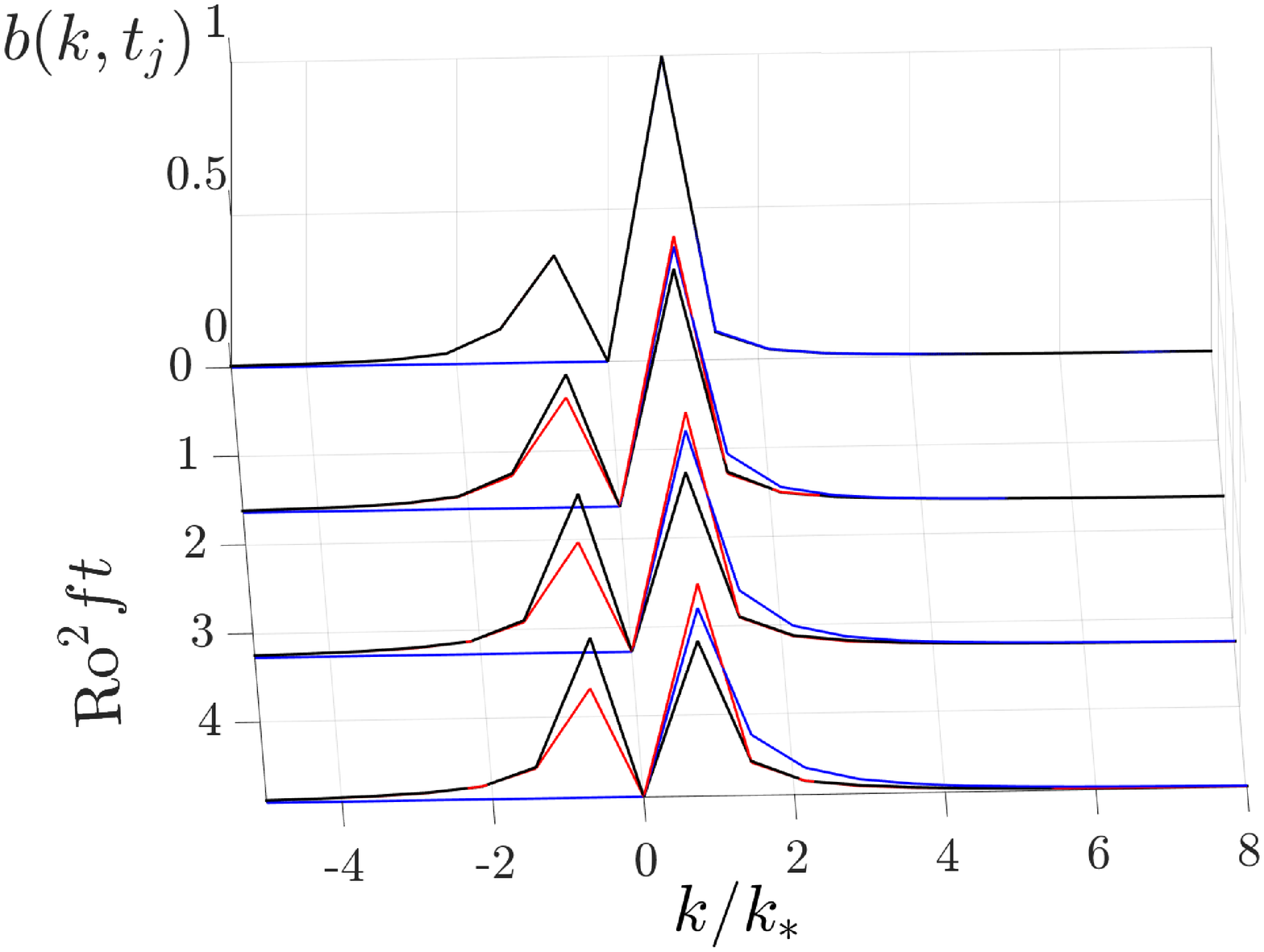} \\
{(c) $\omega=2f$, $\Ro=0.049$, $k_{\mathrm{h}*}/K_{\mathrm{h}*}\simeq 1$} & {(d) $\omega=3f$,  $\Ro=0.099$, $k_{\mathrm{h}*}/K_{\mathrm{h}*}\simeq 1$}
\end{tabular}
\caption{Evolution of the spectrum $b(k,t)$ for IGWs with horizontal wavenumber $k_*$ released in a quasigeostrophic flow with peak wavenumber $K_*$ for the parameters indicated below each panel. Numerical solutions of the Boussinesq equations (black lines) are compared with solutions of the kinetic equation (red lines) and of the diffusion equation (blue lines) that approximates it in the WKBJ limit $k_* \gg K_*$.}
\label{fig:IGW-waterfalls-wkb}
\end{center}
\end{figure}

We first analyse an initial-value problem. Upward-propagating horizontally isotropic IGWs are initialised on the $k_\mathrm{h}=k_\mathrm{h*}$, $k_3=\cot\theta \, k_\mathrm{h*}$, with random phases and an initial kinetic energy $\av{|\bs{u}|^{2}}/2 = 0.1 \av{|\bs{U}|^{2}}/2$. The spectrum $b(k,t)$ at 4 successive times is shown  in Figure \ref{fig:IGW-waterfalls-wkb} for $\omega = 2 f, \, \Ro=0.049$ (left) and $\omega = 3 f, \, \Ro=0.099$ (right), and for $k_\mathrm{h*} / K_\mathrm{h}*  \simeq 4$ (top row) and $k_\mathrm{h*}/ K_\mathrm{h*} \simeq 1$ (bottom row). The results of the Boussinesq simulations are compared with solutions of the kinetic equation and of the diffusion equation of \citet{kafiabad_savva_vanneste_2019}. For the latter two equations, $b(k,t)$ is matched to the spectrum obtained in the Boussinesq simulations after an adjustment period $t_\mathrm{a}\gg(K_*|\vec{c}_g|)^{-1}$, the time for a wavepacket to traverse typical eddies at the IGW group speed, required for the kinetic equation to be valid \citep[][\S5]{besieris,mull-et-al}. The comparison shows a good agreement between the kinetic-equation and Boussinesq results, demonstrating both the ability of the kinetic equation to model faithfully the energy scattering induced by the flow, and the dominance of this process over others such as wave--wave interactions. The diffusion equation provides a good approximation to the spectrum for  $k_\mathrm{h*}/K_\mathrm{h*}  \simeq 4$ but, consistent with its reliance on the assumption $k_* \gg K_*$, is inaccurate $k_\mathrm{h*}/K_\mathrm{h*}  \simeq 1$. 
For the larger $\Ro$ and $k_\mathrm{h*}/K_\mathrm{h*}  \simeq 1$, the match between kinetic-equation and Boussinesq result is poor at low wavenumbers, which could stem from two reasons. First, the discretisation in wavenumber space makes projection onto the constant-frequency cone inaccurate at low wavenumbers, near the cone's apex. Second, the linear wave-vortex decomposition used in this study to extract the wave energy is less accurate around the peak of the geostrophic energy spectrum. As discussed in \cite{kafiabad}, because of the strength of the balanced flow at these scales, a substantial part of what we extract as linear wave modes is in a fact a balanced contribution, `slaved' to the geostrophic modes. A higher-order decomposition would be needed to better isolate the freely propagating waves but is beyond the scope of our study.

\begin{figure}
\begin{center}
 \begin{tabular}{cc}
    \includegraphics[width=.45\linewidth]{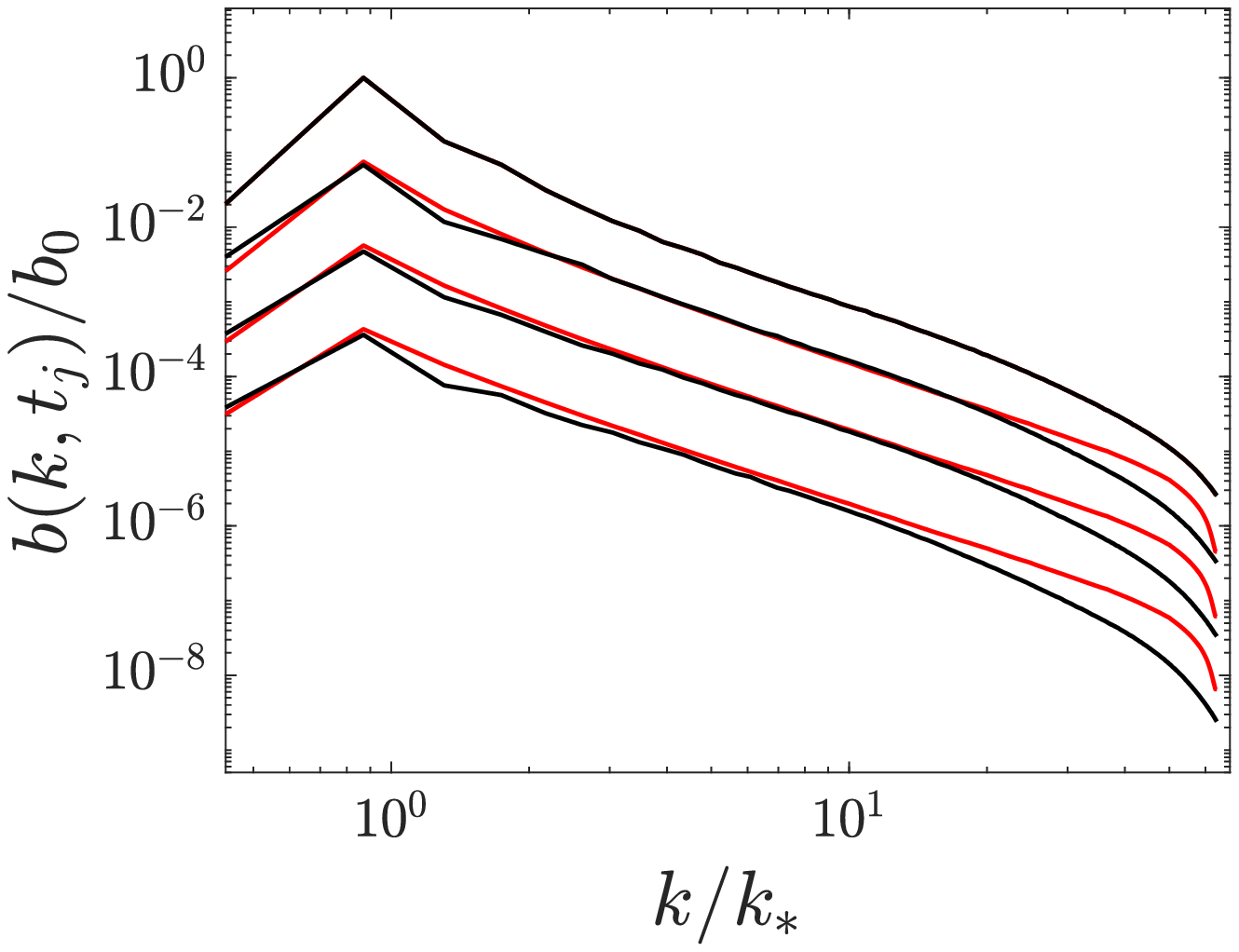} & \includegraphics[width=.45\linewidth]{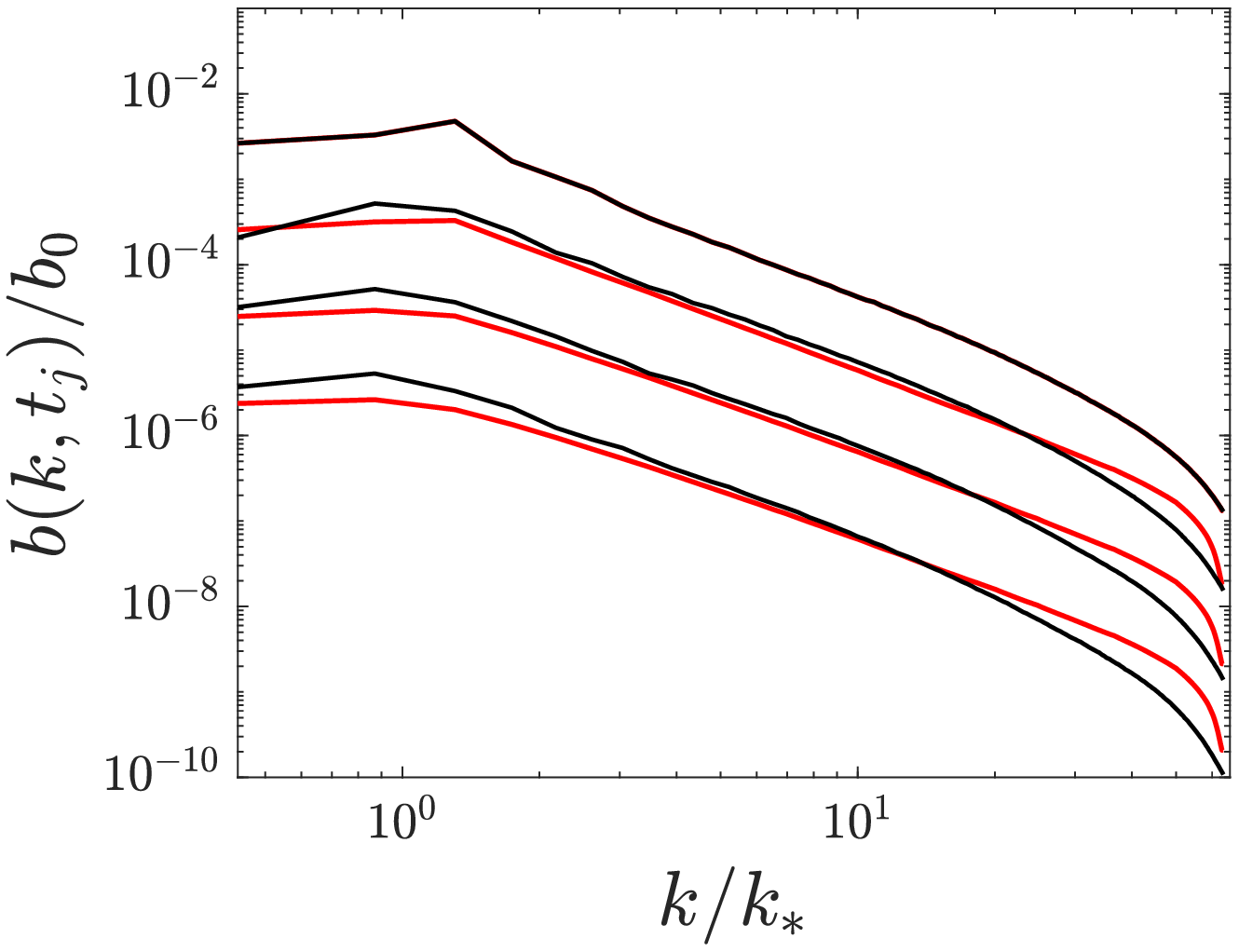} \\
(a) $\omega=2f$, $\Ro=0.049$, upper nappe & (b) $\omega=2f$, $\Ro=0.049$, lower nappe
\end{tabular}
\caption{Log-log representation of the IGW spectrum in Figure \ref{fig:IGW-waterfalls-wkb}c, i.e.\ for $\omega=2f$ and $\Ro=0.049$ obtained from the kinetic equation (red lines) and Boussinesq simulations (black lines); $b_+(k,t)=b(k,t)$, corresponding to the upper nappe of the dispersion-relation cone, is on the left, $b_-(k,t)=b(-k,t)$, corresponding to the lower nappe, is on the right. The curves correspond to the times shown in Figure \ref{fig:IGW-waterfalls-wkb}c and are successively shifted downwards by half a decade for clarity.}

\label{fig:IGW-waterfall-logplot}
\end{center}
\end{figure}

A different view of the results in given by Figure \ref{fig:IGW-waterfall-logplot} which shows the spectrum of upward-propagating waves $b_+(k,t)$ (left) and downward-propagating waves $b_-(k,t)$ for $\omega = 2 f$,  $\Ro=0.049$ and  $k_\mathrm{h*}/ K_\mathrm{h*}  \simeq 1$ in log-log coordinates. This shows an excellent agreement at most but the extreme wavenumbers (where
the dissipation mechanisms, which differ between the kinetic-equation and Boussinesq computations, are felt). Thus the kinetic equation accurately captures the scale cascade that results from scattering by the turbulent flow. Similar results (not shown) are obtained in the WKBJ regime $k_\mathrm{h*} / K_\mathrm{h*} \simeq 4$ where the kinetic equation predicts spectra very close to those obtained in \citet{kafiabad_savva_vanneste_2019} using the diffusion equation. Note that the diffusion equation predicts a $k^{-2} t^{-5}$ dependence of the spectrum which applies for $k \gg K_*$, irrespective of whether the initial wavenumber satisfies the WKBJ condition $k_* \gg K_*$ or not.

\subsection{Forced problem} \label{sec:forced}

\begin{figure}
  \centering
  \includegraphics[width=.5\linewidth]{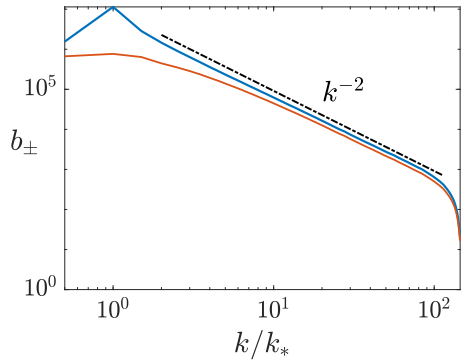}
  \caption{Equilibrium spectra $b_+(k)$ (blue line) and $b_-(k)$ (red line) in a forced solution of the kinetic equation with forcing wavenumber $k_{\mathrm{h}*} \simeq 4K_{h*}$ and $\omega=2f$.}
  \label{fig:IGW-forced-logplot}
\end{figure}

We now turn to a forced problem in which IGWs with random phases are continuously forced along a ring in wavenumber space until they reach a statistically steady state. For the corresponding problem in the WKBJ limit $k_\mathrm{h*} \gg K_\mathrm{h*}$, the forced diffusion equation has an equilibrium power-law solution  $b_\pm(k) \propto k^{-2}$. In general, when the forcing wavenumber is of the order of $ K_\mathrm{h*}$, this power law applies only to the tail of the spectrum; at small and intermediate wavenumbers, the equilibrium spectrum is determined by the steady solution of forced scattering equation
\begin{align}
    {\partial_t \hatt{\vec{b}}_0}&={2\pi}k^2\int_0^\infty {\hatt{\sigmatens}}_{0}(k,k')\,\hatt{\vec{b}}_0({k}',t)\d{k}'    -\Sigma(k)\,\hatt{\vec{b}}_0({k},t)+\vec{\mathcal{F}},\label{eq:forced_iso_kin_equation}
\end{align}
where the forcing term
\begin{equation}
  \vec{\mathcal{F}}=\left( \begin{array}{c}
  A\delta(k-k_*)\\
  0
  \end{array}
   \right),
\end{equation}
with $A$ an arbitrary amplitude, is applied only to upward-propagating waves.  We solve this equation numerically until an approximately steady state is reached
and show the equilibrium spectrum $b_\pm(k)$ obtained in Figure \ref{fig:IGW-forced-logplot}
The parameters chosen are $k_{\mathrm{h}*}\simeq K_{\mathrm{h}*}=4$ for the forcing wavenumber, $\omega=2f$ and $\Ro=0.049$ (note that the equilibrium $b_\pm(k)$ depends only on the shape of the quasigeostrophic-flow spectrum and not on its amplitude). The energy spectrum follows a $k^{-2}$ power law for large $k$, as expected from the WKBJ results of \citet{kafiabad_savva_vanneste_2019}. While the $k^{-2}$ spectrum is an exact stationary solution of the diffusion equation,  for the scattering equation it only holds approximately for $k \gg K_*$. In our setup, the non-diffusive, finite-$k$ effect arise only in a range of wavenumbers close to the forcing wavenumber. Note that \citet{kafiabad_savva_vanneste_2019} confirm the validity of the $k^{-2}$ prediction against Boussinesq solutions and discuss the implications for the interpretation of atmosphere and ocean observations.

Figure \ref{fig:IGW-forced-logplot} shows the spectrum on both the upper and lower nappes of the cones and makes it clear that the stationary spectra of upward- and downward-propagating waves are identical for all wavenumbers outside the immediate vicinity of the forcing wavenumber $k_*$. This is the counterpart for the forced problem to the observation in \S \ref{sec:IVP} that the kinetic equation predicts equipartition of the wave energy between upward and downward-propagating IGWs.

\section{Discussion}

The main result of this paper is the (vector) kinetic equation \eqref{eq:kineq1} governing the energy transfers between upward- and downward-propagating IGWs induced by a turbulent quasigeostrophic flow. The components $\sigma_{\pm}$ of the scattering cross-section tensor, which determine this equation completely, are given in \eqref{sigma_pm}. They depend (linearly) on a single statistic of the quasigeostrophic flow, the kinetic energy spectrum  $\hatt{E}_\mathrm{K}(\vec{k})$.  The main assumption made, that of a small Rossby number, implies that the quasigeostrophic flow evolves slowly enough to be effectively time independent. Accordingly, energy transfers are restricted to IGWs with the same frequency and can be interpreted as resulting from the resonant triadic interactions between two IGWs and a zero-frequency quasigeostrophic (vortical) mode. In wavenumber space, these interactions cause the spreading of IGWs energy along the constant-frequency cones, leading to an isotropisation of the wave energy in the horizontal when the quasigeostrophic flow is horizontally isotropic, and to a cascade to high wavenumbers, that is, to small scales. This cascade depends crucially on the vertical shear of the quasigeostrophic flow and is absent for barotropic flows \citep{savva_vanneste_2018}. In this paper, we focus on the scale cascade by considering the azimuthally-averaged IGW energy spectrum; we leave the study of the process of isotropisation and, in particular, the comparison between its timescale and that of the scale cascade, for future work.

In earlier work \citep{kafiabad_savva_vanneste_2019} we examined IGW scattering in the same setup as here, but with the additional WKBJ assumption of IGW scales much smaller than the typical scale of the quasigeostrophic flow.  Starting from the familiar phase-space transport equation, we derived a diffusion equation for the evolution of IGW energy in wavenumber space. This equation is a limiting form of the kinetic equation derived here, as can be checked directly \citep{savva2020}. In a probabilistic interpretation, the kinetic equation describes a continuous-time random walk, with finite steps in wavenumber space resulting from catalytic interactions, while the diffusion equation describes its Brownian approximation, obtained when in the limit of small steps corresponding to energy transfers that are local in wavenumber space. In this interpretation, the random walk has in fact two states, corresponding the two nappes of the cone or, physically, to IGWs propagating either upwards or downwards.  Transitions between the two states, that is,  transfers between upward- and downward-propagating IGWs are rules out in the WKBJ limit, but are captured by the kinetic equation \eqref{eq:kineq1}.

The results of this paper have potential implications for atmosphere and ocean modelling. As discussed in \citet{kafiabad_savva_vanneste_2019}, the scattering of IGWs by geostrophic turbulence leads to a $k^{-2}$ energy spectrum that is reminiscent of the spectra observed in the atmospheric mesoscale and ocean submesoscale ranges. The results of the present paper make it possible to examine this more fully, by enabling predictions of the IGW statistics across all scale including those that overlap with the geostrophic flow scales. They may also be useful for the parameterisation of IGWs, by providing a quantification of the forward energy flux that results from scattering by unresolved flow. We note that the probabilistic interpretation of the kinetic and diffusion equations mentioned above offers a straightforward route towards stochastic parameterisations of this scattering.

We conclude by pointing out two problems worthy of further study. The first is the relative importance of the scattering by the quasigeostrophic flow and of the nonlinear wave--wave interactions which we have neglected at the outset by linearising the equations of motion. The second concerns the weak energy transfers across constant-frequency cones that stem from the slow time dependence of the flow. Over long timescales, these transfers combine with the along-cone transfers of this paper to yield in a distribution of energy in wavenumber space which could be compared with atmosphere--ocean observations.

\medskip
\noindent
\textbf{Acknowledgments.} HAK and JV are supported by the UK Natural Environment Research Council grant NE/R006652/1. MACS was supported by The Maxwell Institute Graduate School in Analysis and its Applications, a Centre for Doctoral Training funded by the UK Engineering and Physical Sciences Research Council (grant EP/L016508/01), the Scottish Funding Council, Heriot-Watt University and the University of Edinburgh. This work used the ARCHER UK National Supercomputing Service.

\medskip
\noindent
\textbf{Declaration of interests.} The authors report no conflict of interest.

\appendix
\section{Derivation of the kinetic equation} \label{app:kineticderivation}

\subsection{Evolution equation for $\W$}

We start with the Euler--Boussinesq equations written in the form \eqref{eq:IGW_operator_L}, with the linear operator $\L$ defined in \eqref{eq:scaled-equations} and the operator $\NN(\vec{x},\nabla_{\vec{x}})$ grouping the background-flow terms given in terms of its action on $\vec{\phi}=(\gamma,\delta)^{\mathrm{T}}$ by the four components
\begin{subequations} \label{eq:N}
\begin{align}
  \N_{11}\gamma&=(\Omega\nabla)^{-2}\partial_z\Big\{f^2\partial_z\Big([(\psi_x\partial_y-\psi_y\partial_x)-((\nabla_\mathrm{h}^2\psi_x)\partial_y-(\nabla_\mathrm{h}^2\psi_y)\partial_x)\nabla_\mathrm{h}^{-2}]\gamma\Big)\nonumber\\
    &+f^2\nabla_\mathrm{h}^2\Big([(\psi_{xz}\partial_y-\psi_{yz}\partial_x)\nabla_\mathrm{h}^{-2}]\gamma  \Big)+N^2\nabla_\mathrm{h}^2\Big((\psi_x\partial_y-\psi_y\partial_x)\partial_z^{-1}\gamma\Big)  \Big\} ,\label{IGW:N11}\\
  \N_{12}\,\delta&=\nabla^{-2}(f\partial_z)\Big\{\partial_z\Big ([((\nabla_\mathrm{h}^2\psi_x)\partial_x+(\nabla_\mathrm{h}^2\psi_y)\partial_y)\nabla_\mathrm{h}^{-2}+(\nabla_\mathrm{h}^2\psi)\nonumber\\
    &-(\psi_{xz}\partial_x+\psi_{yz}\partial_y+\nabla_\mathrm{h}^2\psi_z)\partial_z^{-1}]\delta\Big)-\nabla_\mathrm{h}^2\Big( [(\psi_{xz}\partial_x+\psi_{yz}\partial_y)\nabla_\mathrm{h}^{-2}-\psi_{zz}\partial_z^{-1}]\delta \Big)\Big\} , \label{IGW:N12} \\
  \N_{21}\gamma&=-2f(\Omega\nabla)^{-2}\partial_{zz}\Big\{  [\psi_{yy}\partial_{xx}-2\psi_{xy}\partial_{xy}+\psi_{xx}\partial_{yy}]\nabla_\mathrm{h}^{-2}\gamma  \Big\} ,\label{IGW:N21}\\
  \N_{22}\,\delta&=\nabla^{-2}\Big\{\partial_{zz}\Big( [(\psi_x\partial_y-\psi_y\partial_x)+2((\psi_{xx}-\psi_{yy})\partial_{xy}-\psi_{xy}(\partial_{xx}-\partial_{yy}))\nabla_\mathrm{h}^{-2}\nonumber\\
    &-(\psi_{xz}\partial_y-\psi_{yz}\partial_x)\partial_z^{-1}]\delta \Big)+\nabla_\mathrm{h}^2\Big( [(\psi_x\partial_y-\psi_y\partial_x)+(\psi_{xz}\partial_y-\psi_{yz}\partial_x)\partial_z^{-1}]\delta \Big)\Big\}.
\end{align}
\end{subequations}
with $\Omega=\Omega(\nabla)$ defined in \eqref{eq:IGW_dispersion_pseudo_operator}.
We derive an evolution equation for the (scaled) Wigner transform of $\vec{\phi}$  differentiating \eqref{eq:scaled-wigner} with respect to $t$ and substituting \eqref{eq:scaled-equations} to obtain
\begin{align}
   & \varepsilon \partial_t \W(\vec{x},\vec{k},t) \\ &={\varepsilon}\int_{\mathbb{R}^3} \e^{\i\vec{k}\cdot\vec{y}}\Big({\partial_t\vec{\phi}(t,\vec{x}-\tfrac{\varepsilon\vec{y}}{2})}{}\vec{\phi}^\mathrm{T}(t,\vec{x}+\tfrac{\varepsilon\vec{y}}{2})+\vec{\phi}(t,\vec{x}-\tfrac{\varepsilon\vec{y}}{2}) {\partial_t\vec{\phi}^\mathrm{T}(t,\vec{x}+\tfrac{\varepsilon\vec{y}}{2})}{} \Big)\frac{\d\vec{y}}{(2\pi)^3} \nonumber \\
    &=-\int_{\mathbb{R}^3} \e^{\i\vec{k}\cdot\vec{y}} \left( \left( \L (\varepsilon\nabla_{\vec{x}})+\eps^{1/2}\NN(\tfrac{\vec{x}}{\varepsilon}-\tfrac{\vec{y}}{2},\varepsilon\nabla_{\vec{x}},\tfrac{t}{\eps^{1/2}})\right) \vec{\phi}(t,\vec{x}-\tfrac{\varepsilon\vec{y}}{2}) \right) \vec{\phi}^\mathrm{T}(t,\vec{x}+\tfrac{\varepsilon\vec{y}}{2})\frac{\d\vec{y}}{(2\pi)^3}+\text{c.c.},
    \label{eq:ho}
\end{align}
where c.c.\ denotes the complex conjugate of the  preceding term. This equation can be closed for $\W(\vec{x},\vec{k},t)$ by introducing the Fourier transform
\begin{equation}
\hatt{\vec{\phi}}(\vec{k},t)=\frac{1}{(2\pi)^3} \int_{\mathbb{R}^3}\e^{\i\vec{k}\cdot\vec{x}}\vec{\phi}(\vec{x},t) \, \d \vec{x}\quad \textrm{and} \quad
   \vec{\phi}(\vec{x},t)=\int_{\mathbb{R}^3}\e^{-\i \vec{k}\cdot\vec{x}}\hatt{\vec{\phi}}(\vec{k},t) \, \d\vec{k}.\label{eq:fourier_convention}
\end{equation}
and noting that the Fourier representation
\begin{equation}
    \W(\vec{x},\vec{k},t)=\varepsilon^{-3}\int_{\mathbb{R}^3}\e^{\i\vec{p}\cdot\vec{x}}\hatt{\vec{\phi}}(- {\vec{k}}/{\varepsilon}-{\vec{p}}/{2},t)\hatt{\vec{\phi}}^*(- {\vec{k}}/{\varepsilon}+{\vec{p}}/{2},t) \, \d\vec{p},
    \label{scaled-duality}
\end{equation}
with $*$ denoting conjugate transpose, can be deduced straightforwardly from \eqref{eq:scaled-wigner} \citep{ryzhik}.
Rewriting $\vec{\phi}$ in terms of $\hatt{\vec{\phi}}$ and  making use of \eqref{scaled-duality}, we rewrite \eqref{eq:ho} as
\begin{multline}
   \varepsilon\partial_t\W(\vec{x},\vec{k},t)+\overbrace{\left(\L(\i\vec{k}+\tfrac{\varepsilon}{2}\nabla_{\vec{x}})\W(\vec{x},\vec{k})+\text{c.c.}\right)}^{\textstyle :=\mathcal{Q}^\varepsilon \W}\\
   +\eps^{1/2}\Bigg(\underbrace{ \int_{\mathbb{R}^3}\e^{-\i\vec{p}\cdot{\boldsymbol\xi}}\NNhat(\vec{p},\i(\vec{k}+\tfrac{\vec{p}}{2})+\tfrac{\varepsilon}{2}\nabla_{\vec{x}},\tau)\W(\vec{x},\vec{k}+\tfrac{\vec{p}}{2}) \, \d\vec{p} +\text{c.c.}}_{\textstyle :=\mathcal{P}^\varepsilon \W} \Bigg)=0,\label{eq:wigner_evolution1}
\end{multline}
where $\vec{\xi}=\vec{x}/\eps$ and $\tau=t/\eps^{1/2}$.
In \eqref{eq:wigner_evolution1} we introduced the matrix $\NNhat$, the Fourier counterpart to the operator $\NN$, defined by the equality
\begin{equation}
 \NN(\vec{x}, \nabla_{\vec{x}})\vec{\phi}(\vec{x}) = \iint \e^{-\i (\vec{q}+\vec{p}) \cdot \vec{x}}\NNhat(\vec{q},-\i\vec{p})\hatt{\vec{\phi}}(\vec{p}) \, \d\vec{q}\d\vec{p}
\end{equation}
holding for all $\vec{\phi}(\vec{x})$. Some care needs to be exercised in deducing the components of $\NNhat$ from those of $\NN$ in \eqref{eq:N} because spatial derivatives act both on the components of $\vec{\phi}(x)$ and on $\psi(\vec{x})$. We note that the components of $ \NN(\vec{x},\nabla_{\vec{x}}) \vec{\phi}$ are sums of the form
\begin{equation}
    \N_{ij}(\vec{x},\nabla_{\vec{x}}) \phi_j(\vec{x})=\sum_k \partial_{\vec{x}}^{\vec{\alpha}}\Big[G_{ij}^k(\vec{x})\partial_{\vec{x}}^{\vec{\beta}}\phi_j(\vec{x})\Big],
\end{equation}
where $\vec{\alpha},\,\vec{\beta}$ are multi-indices (depending on $(i,j,k)$) and $G_{ij}^k(\vec{x})$ depends linearly on $\psi(\vec{x})$.
We then have
\begin{align}
  \N_{ij}(\vec{x}, \nabla_{\vec{x}}) \phi_j &= \sum_k \partial_{\vec{x}}^{\vec{\alpha}}\Bigg[\int_{\mathbb{R}^3} \e^{-\i\vec{q} \cdot \vec{x}}\hatt{G}_{ij}^k (\vec{q})\d\vec{q}\,\partial_{\vec{x}}^{\vec{\beta}}\,\int e^{-\i\vec{p} \cdot \vec{x} }\hatt{\phi}_j(\vec{p})\, \d\vec{p}\Bigg] \nonumber \\
  &= \sum_k \iint_{\mathbb{R}^3} \e^{-\i(\vec{q}+\vec{p}) \cdot \vec{x}}(-\i(\vec{q}+\vec{p}))^{\vec{\alpha}}(-\i\vec{p})^{\vec{\beta}}\hatt{G}_{ij}^k (\vec{q})\hatt{\phi}_j(\vec{p}) \, \d\vec{q}\d\vec{p},
\end{align}
from which we deduce the formula
\begin{equation}
\Nhat_{ij}(\vec{q},-\i\vec{p})  = \sum_k (-\i(\vec{q}+\vec{p}))^{\vec{\alpha}}(-\i\vec{p})^{\vec{\beta}}\hatt{G}_{ij}^k (\vec{q}).
\end{equation}
which makes it possible to compute $\NNhat$ from \eqref{eq:N}. Since $\NNhat$ is a linear function of $\hatt{\psi}$, we can define a matrix $\UUhat(\vec{q},\i\vec{p})$ by
\begin{equation}
\NNhat(\vec{q},\i\vec{p})=\UUhat(\vec{q},\i\vec{p})\hatt{\psi}(\vec{q}).
\label{eq:NU}
\end{equation}
The computation of the cross section below requires this matrix with arguments $\vec{q}=\vec{k}'-\vec{k}$ and $\i \vec{p} = \i \vec{k}'$.
We therefore record the components of $\UUhat(\vec{k}'-\vec{k},\i\vec{k}')$, found to to be
\begin{subequations} \label{UUhat-components}
\begin{align}
    \Uhat_{11}(\vec{k}'-\vec{k},\i\vec{k}')&=\frac{\hat{\vec{k}}_3\cdot\vec{k}_\mathrm{h}'\times\vec{k}_\mathrm{h}}{|\vec{k}_\mathrm{h}'|^2}\frac{|k_3|^2}{|\vec{k}|^2}\Big[\frac{f^2}{\omega^2}(2\vec{k}_\mathrm{h}\cdot\vec{k}_\mathrm{h}'-|\vec{k}_\mathrm{h}||\vec{k}_\mathrm{h}'|\sgn(k_3k_3'))+\frac{N^2}{\omega^2}\frac{|\vec{k}_\mathrm{h}|^2|\vec{k}_\mathrm{h}'|^2}{k_3k_3'} \Big], \\
    \Uhat_{22}(\vec{k}'-\vec{k},\i\vec{k}')&=\frac{\hat{\vec{k}}_3\cdot\vec{k}_\mathrm{h}'\times\vec{k}_\mathrm{h}}{|\vec{k}_\mathrm{h}'|^2}\frac{|k_3|^2}{|\vec{k}|^2}\Big[2\vec{k}_\mathrm{h}\cdot\vec{k}_\mathrm{h}'-|\vec{k}_\mathrm{h}||\vec{k}_\mathrm{h}'|\sgn(k_3k_3')+\frac{|\vec{k}_\mathrm{h}|^2|\vec{k}_\mathrm{h}'|^2}{k_3k_3'}\Big],  \\
    \Uhat_{12}(\vec{k}'-\vec{k},\i\vec{k}')&=\frac{f}{|\vec{k}_\mathrm{h}'|^2}\frac{|k_3|^2}{|\vec{k}|^2}\Big[(k_3'-k_3)^2\frac{|\vec{k}_\mathrm{h}'||\vec{k}_\mathrm{h}|}{|k_3'||k_3|}-|\vec{k}_\mathrm{h}'-\vec{k}_\mathrm{h}|^2  \Big]  \vec{k}_\mathrm{h}'\cdot\vec{k}_\mathrm{h},  \\
    \Uhat_{21}(\vec{k}'-\vec{k},\i\vec{k}')&=2\frac{f}{\omega^2}\frac{|k_3|^2}{|\vec{k}|^2}\frac{|\vec{k}_\mathrm{h}'\times\vec{k}_\mathrm{h}|^2}{|\vec{k}_\mathrm{h}'|^2},
\end{align}
\end{subequations}
after a lengthy calculation that uses $\Omega(\i\vec{k})=\omega(\vec{k})$.

\subsection{Multiscale asymptotics}

We now derive the asymptotic limit of \eqref{eq:wigner_evolution1} using a multiscale expansion. We introduce the expansion
\eqref{wigner_multiscale} into \eqref{eq:wigner_evolution1}, expanding
the differential operators as
\begin{equation}
    \nabla_{\vec{x}}\mapsto\nabla_{\vec{x}}+\varepsilon^{-1}\nabla_{\boldsymbol\xi} \quad \textrm{and} \quad \partial_t\mapsto\partial_t+\varepsilon^{-1/2}\partial_\tau,
\end{equation}
where $\vec{x}$ and $\boldsymbol\xi$, $t$ and $\tau$ are  treated as independent variables, leading to the expansion
\begin{align}
    \mathcal{Q}^\varepsilon=\mathcal{Q}_0+\varepsilon\mathcal{Q}_1+O(\varepsilon^2),\;\;\;\mathcal{P}^\varepsilon=\mathcal{P}_0+\varepsilon\mathcal{P}_1+{O}(\varepsilon^2)
    \label{aa}
\end{align}
of the operators in \eqref{eq:wigner_evolution1}.
It turns out that only the leading order term $\mathcal{P}_0$ is required for the derivation of the kinetic equation.

The operators in \eqref{aa} can be written explicitly through their action on an arbitrary function $Z(\vec{x},\vec{\xi},\vec{k})$:
\begin{align}
    \widetilde{\mathcal{Q}}_0Z(\vec{x},\vec{\xi},\vec{k})&= \L(\i \vec{k}+\tfrac{1}{2}\nabla_{\boldsymbol\xi})Z(\vec{x},\vec{\xi},\vec{k})+\text{c.c.} \label{Q_0} \\
    \widetilde{\mathcal{Q}}_1Z(\vec{x},\vec{\xi},\vec{k})&=\frac{1}{2\i{}}\big[\nabla_{\vec{k}}\L(\i \vec{k}+\tfrac{1}{2}\nabla_{\boldsymbol\xi})\big] \cdot\nabla_{\vec{x}}Z(\vec{x},\vec{\xi},\vec{k})  +\text{c.c.}\label{Q_1}\\
    \widetilde{\mathcal{P}}_0Z(\vec{x},\vec{\xi},\vec{k})&=\int_{\mathbb{R}^3}\e^{-\i \vec{p}\cdot\boldsymbol\xi}\NNhat(\vec{p},\i (\vec{k}+\tfrac{\vec{p}}{2})+\tfrac{1}{2}\nabla_{\boldsymbol\xi},\tau)Z(\vec{x},\vec{\xi},\vec{k}+\tfrac{\vec{p}}{2})\d\vec{p}  +\text{c.c.}\label{P_0}
\end{align}
We have decorated the operators with a tilde to highlight the presence of $\nabla_{\vec{\xi}}$ in their definition; the tildes will be removed whenever this dependence disappears.

Substituting the operators into \eqref{eq:wigner_evolution1} gives us the evolution equation for the Wigner function as
\begin{equation}
    \Big[\frac{1}{\varepsilon}\widetilde{\mathcal{Q}}_0+\frac{1}{\eps^{1/2}}\Big(\widetilde{\mathcal{P}}_0+\frac{\partial}{\partial\tau}\Big)+\Big(\widetilde{\mathcal{Q}}_1+\frac{\partial}{\partial t}\Big)  \Big]\W^\varepsilon(\vec{x},\boldsymbol\xi,\vec{k},t,\tau)=0.\label{Wigner_evolution}
\end{equation}
Introducing  the expansion \eqref{wigner_multiscale} then leads to a hierarchy of equations to be solved at each order in $\varepsilon$.

The leading-order equation is
\begin{equation}
    \mathcal{Q}_0\W^{(0)}=\L(\i\vec{k})\W^{(0)}(\vec{x},\vec{k},t)+\text{c.c.}=0\label{Oe-1}
\end{equation}
whose general solution
\begin{equation}
    \W^{(0)}(\vec{x},\vec{k},t)=\sum_{j=\pm}a_j(\vec{x},\vec{k},t)\E_j(\vec{k})\label{W0},
\end{equation}
is a linear combination of the matrices $\E_j(\vec{k})=\vec{e}_j(\vec{k})\vec{e}^*_j(\vec{k})$ constructed from the (right) eigenvectors $\vec{e}_j(\vec{k})$ of $ \L(\i\vec{k})$ (see \eqref{eq:evectors}).
The so-far undetermined amplitudes  $a_j(\vec{x},\vec{k},t)$ are real because the Wigner function is Hermitian.

At $O(\varepsilon^{-1/2})$, we find
\begin{equation}
    \widetilde{\mathcal{Q}}_0\W^{(1)}(\vec{x},\boldsymbol\xi,\vec{k},t,\tau)=-\mathcal{P}_0 \W^{(0)}(\vec{x},\vec{k},t),\label{Oe-1/2}
\end{equation}
where we have used that $\partial_\tau \W^{(0)}=0$.
To solve \eqref{Oe-1/2}, we rewrite $\W^{(1)}$ in terms of its Fourier transform with respect to $\boldsymbol\xi$,
\begin{equation}
    \W^{(1)}(\vec{x},\boldsymbol\xi,\vec{k},t,\tau)=\int_{\mathbb{R}^3}\e^{-\i\vec{p}\cdot\boldsymbol\xi}\What^{(1)}(\vec{x},\vec{p},\vec{k},t,\tau)\d\vec{p}.
\end{equation}
Substituting this into \eqref{Oe-1/2} yields
\begin{multline}  \label{eq:W1}
   \L(\i(\vec{k}-\tfrac{\vec{p}}{2}))\What^{(1)}(\vec{p},\vec{k})+\Big[\L(\i(\vec{k}+\tfrac{\vec{p}}{2}))\What^{(1)}(-\vec{p},\vec{k})\Big]^* + \theta\What^{(1)}(\vec{p},\vec{k})\\
   =-\NNhat\big(\vec{p},\i(\vec{k}+\tfrac{\vec{p}}{2})\big)\W^{(0)}(\vec{k}+\tfrac{\vec{p}}{2})-\Big[\NNhat\big(-\vec{p},\i(\vec{k}-\tfrac{\vec{p}}{2})\big)\W^{(0)}(\vec{k}-\tfrac{\vec{p}}{2})\Big]^*,
\end{multline}
where we have suppressed dependencies on $\vec{x}$, $t$ and $\tau$ for conciseness. Following \cite{ryzhik}, we have introduced a regularisation parameter $\theta>0$  which will be taken to zero at a later stage.

We solve \eqref{eq:W1} by projection on the left eigenvectors of $\L(\i\vec{k})$, that is, on the row vectors $\vec{c}_j$ solving
\begin{equation}
  \vec{c}_j\L=\i\omega_j\vec{c}_j.
  \label{left_eigenvectors2}
\end{equation}
and satisfying
\begin{equation}
\vec{c}_j=\vec{e}_j^*\M\quad\textrm{and}\quad\vec{c}_i\vec{e}_j=\delta_{ij}
\end{equation}
as can be shown using that $\M(\vec{k}) \L(\i \vec{k})$ is skew-Hermitian.  Pre- and post-multiplying \eqref{eq:W1} by $\vec{c}_n(\vec{k}-\vec{p}/2)$ and $\vec{c}^*_m(\vec{k}+\vec{p}/2)$ and using that  $\What^{(1)}(\vec{p},\vec{k})=\What^{(1)*}(-\vec{p},\vec{k})$ (because the the Wigner transform is Hermitian) gives
\begin{align}
&   -\Big(\i(\omega_n(\vec{k}-\tfrac{\vec{p}}{2})-\omega_m(\vec{k}+\tfrac{\vec{p}}{2}))+\theta\Big)\vec{c}_n(\vec{k}-\tfrac{\vec{p}}{2})\What^{(1)}(\vec{p},\vec{k})\vec{c}_m^*(\vec{k}+\tfrac{\vec{p}}{2}) \nonumber \\
   =&\sum_{i=\pm}a_i(\vec{k}+\tfrac{\vec{p}}{2})\vec{c}_n(\vec{k}-\tfrac{\vec{p}}{2})\NNhat(\vec{p},\i(\vec{k}+\tfrac{\vec{p}}{2}))\vec{e}_i(\vec{k}+\tfrac{\vec{p}}{2})\vec{e}_i^*(\vec{k}+\tfrac{\vec{p}}{2})\vec{c}_m^*(\vec{k}+\tfrac{\vec{p}}{2}) \nonumber \\
   +&\sum_{j=\pm}a_j(\vec{k}-\tfrac{\vec{p}}{2})\vec{c}_n(\vec{k}-\tfrac{\vec{p}}{2})\vec{e}_j(\vec{k}-\tfrac{\vec{p}}{2})\vec{e}_j^*(\vec{k}-\tfrac{\vec{p}}{2})\NNhat^*(-\vec{p},\i(\vec{k}-\tfrac{\vec{p}}{2}))\vec{c}_m^*(\vec{k}+\tfrac{\vec{p}}{2}).
\end{align}
We now decompose $\What^{(1)}$ using the vectors $\vec{e}_i(\vec{k})$, which form a complete basis, as
\begin{equation}
    \What^{(1)}(\vec{x},\vec{p},\vec{k},t,\tau)=\sum_{m,n=\pm}\alpha_{mn}(\vec{x},\vec{p},\vec{k},t,\tau)\vec{e}_n(\vec{k}-\tfrac{\vec{p}}{2})\vec{e}_m^*(\vec{k}+\tfrac{\vec{p}}{2}).
\end{equation}
Using this along with the orthonormality of the eigenvectors and \eqref{eq:NU} we finally write the solution
\begin{multline}
    \What^{(1)}(\vec{x},\vec{p},\vec{k},t,\tau)=\sum_{m,n=\pm}\Big[a_m(\vec{x},\vec{k}+\tfrac{\vec{p}}{2},t)\vec{c}_n(\vec{k}-\tfrac{\vec{p}}{2})\UUhat(\vec{p},\i(\vec{k}+\tfrac{\vec{p}}{2}))\vec{e}_m(\vec{k}+\tfrac{\vec{p}}{2})\\
    +a_n(\vec{x},\vec{k}-\tfrac{\vec{p}}{2},t)\vec{e}_n^*(\vec{k}-\tfrac{\vec{p}}{2})\UUhat^*(-\vec{p},\i(\vec{k}-\tfrac{\vec{p}}{2}))\vec{c}^*_m(\vec{k}+\tfrac{\vec{p}}{2})\Big]\frac{\vec{e}_n(\vec{k}-\tfrac{\vec{p}}{2})\vec{e}_m^*(\vec{k}+\tfrac{\vec{p}}{2})\hatt{\psi}(\vec{p},\tau)}{\i\big(\omega_m(\vec{k}+\tfrac{\vec{p}}{2})-\omega_n(\vec{k}-\tfrac{\vec{p}}{2})\big)-\theta},
\end{multline}
where we have taken into account that $\hatt{\psi }(\vec{p})=\hatt{\psi }^*(-\vec{p})$.
We note that this solution shows $\W^{(1)}$ is linear in the random field $\psi$.

The slow evolution of the leading-order Wigner function $\W^{(0)}$ is controlled by the $O(1)$ term in the expansion of  \eqref{Wigner_evolution}, given by
\begin{equation}
    -\widetilde{\mathcal{Q}}_0\W^{(2)}=(\widetilde{\mathcal{P}}_0+\partial_\tau)\W^{(1)}+(\mathcal{Q}_1+\partial_t)\W^{(0)}.\label{Oe0}
\end{equation}
We assume that the random streamfunction is a stationary process in $\tau$ and homogeneous in $\vec{\xi}$, with zero mean, $\avg{\psi(\boldsymbol\xi,\tau)}=0$, and covariance
\begin{equation}
    \avg{\psi(\boldsymbol\xi,\tau)\psi(\boldsymbol\xi',\tau)}=R(\boldsymbol\xi-\boldsymbol\xi'),
\end{equation}
where $\avg{\cdot}$ denotes an ensemble average, or equivalently an average over $\boldsymbol\xi$. In terms of Fourier transforms, this implies that
\begin{equation}
    \avg{\hatt{\psi}(\vec{p})\hatt{\psi}(\vec{p}')}=\hatt{R}(\vec{p})\delta(\vec{p}+\vec{p}')\label{power_spectrum},
\end{equation}
where the streamfunction power spectrum $\hatt{R}$ is the Fourier transform of $R$. Then since $\vec{u}=\nabla_\mathrm{h}^\bot\psi$, the more familiar kinetic energy spectrum is then
\begin{equation}
\hatt{E}_\mathrm{K}(\vec{k})=\abs{\vec{k}_\mathrm{h}}^2\hatt{R}(\vec{k}).\label{energy_spectrum}
\end{equation}

We now take the average of \eqref{Oe0}.
 The slow time derivative term on the right-hand side disappears since $\avg{\W^{(1)}}=0$. Since $\avg{\partial_\xi \W^{(2)}}=0$,  $\avg{\widetilde{\mathcal{Q}}_0 \W^{(2)}}=\mathcal{Q}_0 \avg{\W^{(2)}}$, where the removal of the tilde corresponds to setting $\nabla_{\vec\xi}$ to $0$ in $\mathcal{Q}_0$. This leads to
\begin{equation}
    -{\mathcal{Q}}_0\W^{(2)}=\avg*{\widetilde{\mathcal{P}}_0\W^{(1)}+(\mathcal{Q}_1+\partial_t)\W^{(0)}},\label{Oe0_averaged}
\end{equation}
an inhomogeneous version of \eqref{Oe-1}.

The matrix $\mathcal{Q}_0$ has a non-trivial null space, spanned by the matrices $\E_j(\vec{k})$;  the right-hand side of \eqref{Oe0_averaged} must therefore satisfy a solvability condition. Since  $\i\mathcal{Q}_0 = \i \L(\i \vec{k})$ is self-adjoint with respect to the matrix inner product
\begin{equation}
    \minner{\XX}{\YY}:=\text{tr}(\M\XX^*\M\YY),
\end{equation}
this condition is obtained by applying $\minner{\E_j(\vec{k})}{\cdot}$ to \eqref{Oe0_averaged}. We deal with the resulting terms one by one. First, by orthogonality and \eqref{W0} we have
\begin{equation}
    \minner{\E_i}{\partial_t\W^{(0)}}=\sum_{j=\pm}(\partial_ta_j)\minner{\E_i}{\E_j}=\partial_ta_i(\vec{x},\vec{k},t).\label{eq:partial_ta_i}
\end{equation}
Next,
\begin{align}
   \minner{\E_i}{\mathcal{Q}_1\W^{(0)}}&=\sum_{j=\pm}\frac{1}{2 \i{}}\minner{\E_i}{(\nabla_{\vec{k}}\L\cdot\nabla_{\vec{x}}a_j)\E_j}+\text{c.c.} \nonumber \\
   &=\sum_{j=\pm}\frac{1}{2\i{}}\minner{\E_i}{\nabla_{\vec{k}}(\L\E_j)-\L\nabla_{\vec{k}}\E_j}\cdot\nabla_{\vec{x}}a_j+\text{c.c.} \nonumber \\
   &=\sum_{j=\pm}\frac{1}{2\i{} }\minner{\E_i}{\nabla_{\vec{k}}(\i\omega_j)\E_j+(\i\omega_j-\L)\nabla_{\vec{k}}\E_j}\cdot\nabla_{\vec{x}}a_j+\text{c.c.} \nonumber \\
   &=\nabla_{\vec{k}}\omega_i\cdot \nabla_{\vec{x}}a_i(\vec{x},\vec{k},t).\label{eq:advec_a_i}
\end{align}

In order to evaluate the remaining term, we note that, using \eqref{eq:NU} and \eqref{power_spectrum}, we have
\begin{equation}
    \avg*{\Nhat_{\alpha\beta}(\vec{p},\i\vec{q})\Nhat_{\gamma\delta}(\vec{p}',\i\vec{q}')}=\Uhat_{\alpha\beta}(\vec{p},\i\vec{q})\Uhat_{\gamma\delta}(\vec{p}',\i\vec{q}')\hatt{R}(\vec{p})\delta(\vec{p}+\vec{p}'),\label{VV}
\end{equation}
where Greek indices are used for matrix elements to make the following derivation clearer, and  summation over repeated Greek indices is implied.

Expanding all terms, and making use of the delta function in \eqref{VV}, we have
\begin{align}
  &  \minner{\E_i}{\avg*{\widetilde{\mathcal{P}}_0\W^{(1)}}} \nonumber\\
  &=\iint\e^{\i(\vec{p}+\vec{p}')\cdot\boldsymbol\xi}\MM_{\nu\rho}e_\rho^i(\vec{k})e_\sigma^{i*}(\vec{k})\MM_{\sigma\lambda}\avg*{\Nhat_{\lambda\mu}\big(\vec{p},\i(\vec{k}+\tfrac{\vec{p}-\vec{p}'}{2})\big)\What^{(1)}_{\mu\nu}(\vec{p}',\vec{k}+\tfrac{\vec{p}}{2})}\d\vec{p}\d\vec{p}'+\text{c.c.} \nonumber \\
   &=\int\sum_{m,n=\pm} c^i_\lambda(\vec{k})\Uhat_{\lambda\mu}(\vec{p},\i(\vec{k}+\vec{p}))e_\mu^n(\vec{k}+\vec{p})\overbrace{c_\rho^m(\vec{k})e^i_\rho(\vec{k})}^{\delta^{im}}\hatt{R}(\vec{p}) \nonumber \\
   &\hspace{0em}\times  \frac{a_m(\vec{k})c_\alpha^n(\vec{k}+\vec{p})\Uhat_{\alpha\beta}(-{\vec{p}},\i\vec{k})e_\beta^m(\vec{k})+a_n(\vec{k}+\vec{p})\Big(c_\alpha^m(\vec{k})\Uhat_{\alpha\beta}({\vec{p}},\i(\vec{k}+\vec{p}))e_\beta^n(\vec{k}+\vec{p})\Big)^*}{\i\big(\omega_m(\vec{k})-\omega_n(\vec{k}+\vec{p})\big)-\theta} \d\vec{p} \nonumber \\
   &\hspace{1em}+\text{c.c.} \nonumber \\
   &=-2\theta\Re\int\sum_{n=\pm}c_\lambda^i(\vec{k})\Uhat_{\lambda\mu}({\vec{k}'-\vec{k}},\i\vec{k}')e_\mu^n(\vec{k}')\hatt{R}(\vec{k}'-\vec{k}) \nonumber \\
   &\hspace{1em}\times\frac{a_i(\vec{k})c_\alpha^n(\vec{k}')\Uhat_{\alpha\beta}(\vec{k}-\vec{k}',\i\vec{k})e^i_\beta(\vec{k})+a_n(\vec{k}')\Big(c_\alpha^i(\vec{k})\Uhat_{\alpha\beta}(\vec{k}'-\vec{k},\i\vec{k}')e^n_\beta(\vec{k}')\Big)^*}{\big(\omega_i(\vec{k})-\omega_n(\vec{k}')\big)^2+\theta^2}\d\vec{k}',\label{eq:large-expression}
\end{align}
where we have let $\vec{k}':=\vec{k}+\vec{p}$. Setting the regularisation parameter $\theta\to0$, we have that $\theta/(x^2+\theta^2)\to\pi\delta(x)$. This leads to a factor $\delta(\omega_i(\vec{k})-\omega_n(\vec{k}'))$ which indicates that scattering is restricted within a single branch of the dispersion relation, and so we may drop the sum over $n$ and let $i=n$.

We simplify \eqref{eq:large-expression} by computing
\begin{subequations} \label{eq:igw_alpha_beta}
\begin{align}
 c^\pm_\lambda(\vec{k})\Uhat_{\lambda\mu}(\vec{k}'-\vec{k},\i\vec{k}')b^\pm_\mu(\vec{k}') &= \inner{\vec{e}_\pm(\vec{k})}{\UUhat(\vec{k}'-\vec{k},\i\vec{k}')\vec{e}_\pm(\vec{k}')}_\M \\
 &= \frac{1}{2\omega}\frac{|\vec{k}||\vec{k}_\mathrm{h}'||k_3'|}{|\vec{k}'||\vec{k}_\mathrm{h}||k_3|}[\omega(\Uhat_{11}+\Uhat_{22})\pm \i(\Uhat_{12}-\omega^2\Uhat_{21})] \\
 &=:\alpha(\vec{k},\vec{k}')\pm\i\beta(\vec{k},\vec{k}'),
\end{align}
\end{subequations}
using  \eqref{igw_matrix_M} and \eqref{UUhat-components} and omitting the arguments $(\vec{k}'-\vec{k},\i\vec{k}')$ of the functions $\Uhat_{ij}$. The last line defines the two real functions $\alpha(\vec{k},\vec{k}')$ and $\beta(\vec{k},\vec{k}')$ which can be written down using the explicit expressions for $\Uhat_{ij}$ in \eqref{UUhat-components}. The symmetry properties
\begin{equation}
    \alpha(\vec{k},\vec{k}')=-\alpha(\vec{k}',\vec{k}) \;\;\;\text{and}\;\;\; \beta(\vec{k},\vec{k}')=\beta(\vec{k}',\vec{k})\label{eq:al-bet-symmetry}
\end{equation}
can be verified from these expressions. Using \eqref{eq:igw_alpha_beta}--\eqref{eq:al-bet-symmetry}, the terms in \eqref{eq:large-expression} simplify as
\begin{align}
   \Re\; \Big(c_\lambda(\vec{k})\Uhat_{\lambda\mu}(\vec{k}'-\vec{k},\i\vec{k}')e_\mu(\vec{k}')\Big)\Big(c_\alpha(\vec{k}')\Uhat_{\alpha\beta}(\vec{k}-\vec{k}',\i\vec{k})e_\beta(\vec{k})\Big) &=-(\alpha^2(\vec{k},\vec{k}')+\beta^2(\vec{k},\vec{k}')), \\
   \Re\; \Big(c_\lambda(\vec{k})\Uhat_{\lambda\mu}(\vec{k}'-\vec{k},\i\vec{k}')e_\mu(\vec{k}')\Big)\Big(c_\alpha(\vec{k})\Uhat_{\alpha\beta}(\vec{k}'-\vec{k},\i\vec{k}')e_\beta(\vec{k}')\Big)^* &=\alpha^2(\vec{k},\vec{k}')+\beta^2(\vec{k},\vec{k}'),
\end{align}
and \eqref{eq:large-expression} simplifies to
\begin{equation}
     \minner{\E_i}{\avg*{\widetilde{\mathcal{P}}_0\W^{(1)}}} \label{scat-term}\\=
     2\pi\int_{\mathbb{R}^3}\big(\alpha^2(\vec{k},\vec{k}') +\beta^2(\vec{k},\vec{k}') \big)\hatt{R}(\vec{k}'-\vec{k})\delta\big(\omega(\vec{k})-\omega(\vec{k}')\big)\big[a(\vec{k})-a(\vec{k}')\big]\d\vec{k}'.
\end{equation}
Combining this result with \eqref{eq:partial_ta_i} and \eqref{eq:advec_a_i} reduces the solvability condition for \eqref{Oe0_averaged}
to the kinetic equation \eqref{eq:kineq_full}, with the cross section
\begin{equation}
    \sigma(\vec{k},\vec{k}'):=2\pi\big(\alpha^2(\vec{k},\vec{k}')+\beta^2(\vec{k},\vec{k}')\big)\hatt{R}(\vec{k}'-\vec{k})\delta\big(\omega(\vec{k})-\omega(\vec{k}')\big).\label{eq:general-cross-section}
\end{equation}
Replacing the streamfunction spectrum $\hatt{R}$ by the kinetic-energy spectrum $\hatt{E}_\mathrm{K}$ using \eqref{power_spectrum} and substituting explicit expressions for $\alpha(\vec{k},\vec{k}')$ and $\beta(\vec{k},\vec{k}')$ gives the full form \eqref{eq:IGW-full-cross-section} of the cross section.

\bibliographystyle{jfm}
\bibliography{mybib}

\begin{thebibliography}{43}
\expandafter\ifx\csname natexlab\endcsname\relax\def\natexlab#1{#1}\fi
\def\au#1{#1} \def\ed#1{#1} \def\yr#1{#1}\def\at#1{#1}\def\jt#1{\textit{#1}}
  \def\bt#1{#1}\def\bvol#1{\textbf{#1}} \def\vol#1{#1} \def\pg#1{#1}
  \def\publ#1{#1}\def\arxiv#1{#1}\def\org#1{#1}\def\st#1{\textit{#1}}

\bibitem[Andrews \& McIntyre(1978)]{andrews1978exact}
{\sc \au{Andrews, D.~G.} \& \au{McIntyre, M.~E.}} \yr{1978}  \at{An exact
  theory of nonlinear waves on a {L}agrangian-mean flow}.  \jt{J. Fluid Mech.}
  \bvol{89},  \pg{609--646}.

\bibitem[Asselin {\em et~al.\/}(2018)Asselin, Bartello \& Straub]{asselin2018}
{\sc \au{Asselin, O.}, \au{Bartello, P.} \& \au{Straub, D.~N.}} \yr{2018}
  \at{On boussinesq dynamics near the tropopause}.  \jt{J. Atmos. Sci.}
  \bvol{75}~(2),  \pg{571--585}.

\bibitem[Bal {\em et~al.\/}(2010)Bal, Komorowski \& Ryzhik]{bal-et-al}
{\sc \au{Bal, G.}, \au{Komorowski, T.} \& \au{Ryzhik, L.}} \yr{2010}
  \at{Kinetic limits for waves in a random medium}.  \jt{Kinetic Rel. Models}
  \bvol{3},  \pg{529--644}.

\bibitem[Bartello(1995)]{bartello}
{\sc \au{Bartello, P.}} \yr{1995}  \at{Geostrophic adjustment and inverse
  cascades in rotating stratified turbulence}.  \jt{J. Atmos. Sci.}
  \bvol{52}~(24),  \pg{4410--4428}.

\bibitem[Besieris(1987)]{besieris}
{\sc \au{Besieris, I.~M.}} \yr{1987}  \at{Stochastic wave-kinetic theory of
  radiative transfer in the presence of ionization}.  \jt{Rad. Sci.}
  \bvol{22}~(6),  \pg{885--888}.

\bibitem[B{\"u}hler(2014)]{buhler_2014}
{\sc \au{B{\"u}hler, O.}} \yr{2014} {\em Waves and Mean Flows\/}, 2nd edn.
  \publ{Cambridge University Press}.

\bibitem[B{\"{u}}hler {\em et~al.\/}(2014)B{\"{u}}hler, Callies \&
  Ferrari]{buhl-et-al14}
{\sc \au{B{\"{u}}hler, O.}, \au{Callies, J.} \& \au{Ferrari, R.}} \yr{2014}
  \at{{Wave-vortex decomposition of one-dimensional ship-track data}}.  \jt{J.
  Fluid. Mech.}  \bvol{756},  \pg{1007--1026}.

\bibitem[B{\"u}hler {\em et~al.\/}(2017)B{\"u}hler, Kuang \&
  Tabak]{buhler_kuang_tabak_2017}
{\sc \au{B{\"u}hler, O.}, \au{Kuang, M.} \& \au{Tabak, E.~G.}} \yr{2017}
  \at{Anisotropic {H}elmholtz and wave–vortex decomposition of
  one-dimensional spectra}.  \jt{J. Fluid Mech.}  \bvol{815},  \pg{361–387}.

\bibitem[Callies {\em et~al.\/}(2016)Callies, Bühler \& Ferrari]{cali-et-al16}
{\sc \au{Callies, J.}, \au{Bühler, O.} \& \au{Ferrari, R.}} \yr{2016}  \at{The
  dynamics of mesoscale winds in the upper troposphere and lower stratosphere}.
   \jt{J. Atmos. Sci.}  \bvol{73}~(12),  \pg{4853--4872}.

\bibitem[Callies \& Ferrari(2013)]{cali-ferr}
{\sc \au{Callies, J.} \& \au{Ferrari, R.}} \yr{2013}  \at{Interpreting energy
  and tracer spectra of upper-ocean turbulence in the submesoscale range
  (1–200 km)}.  \jt{J. Phys. Oceanogr.}  \bvol{43}~(11),  \pg{2456--2474}.

\bibitem[Callies {\em et~al.\/}(2014)Callies, Ferrari \&
  B{\"{u}}hler]{cali-et-al14}
{\sc \au{Callies, J.}, \au{Ferrari, R.} \& \au{B{\"{u}}hler, O.}} \yr{2014}
  \at{{Transition from geostrophic turbulence to inertia–gravity waves in the
  atmospheric energy spectrum}}.  \jt{Proc. Natl. Acad. Sci.}  \bvol{111}~(48),
   \pg{17033--17038}.

\bibitem[Danioux \& Vanneste(2016)]{danioux}
{\sc \au{Danioux, E.} \& \au{Vanneste, J.}} \yr{2016}  \at{Near-inertial-wave
  scattering by random flows}.  \jt{Phys. Rev. Fluids}  \bvol{1},  \pg{033701}.

\bibitem[Eden {\em et~al.\/}(2019)Eden, Chouksey \&
  Olbers]{eden-chouksey-olbers-2019}
{\sc \au{Eden, C.}, \au{Chouksey, M.} \& \au{Olbers, D.}} \yr{2019}  \at{Mixed
  rossby–gravity wave–wave interactions}.  \jt{J. Phys. Oceanogr.}
  \bvol{49},  \pg{291--308}.

\bibitem[Gilbert \& Vanneste(2018)]{gilbert2018geometric}
{\sc \au{Gilbert, A.~D.} \& \au{Vanneste, J.}} \yr{2018}  \at{Geometric
  generalised {L}agrangian-mean theories}.  \jt{J. Fluid Mech.}  \bvol{839},
  \pg{95--134}.

\bibitem[Kafiabad \& Bartello(2016)]{kafiabad}
{\sc \au{Kafiabad, H.~A.} \& \au{Bartello, P.}} \yr{2016}  \at{Balance dynamics
  in rotating stratified turbulence}.  \jt{J. Fluid Mech.}  \bvol{795},
  \pg{914–949}.

\bibitem[Kafiabad \& Bartello(2018)]{kafi-bart}
{\sc \au{Kafiabad, Hossein~A} \& \au{Bartello, Peter}} \yr{2018}
  \at{Spontaneous imbalance in the non-hydrostatic boussinesq equations}.
  \jt{Journal of Fluid Mechanics}  \bvol{847},  \pg{614--643}.

\bibitem[Kafiabad {\em et~al.\/}(2019)Kafiabad, Savva \&
  Vanneste]{kafiabad_savva_vanneste_2019}
{\sc \au{Kafiabad, H.~A.}, \au{Savva, M. A.~C.} \& \au{Vanneste, J.}} \yr{2019}
   \at{Diffusion of inertia-gravity waves by geostrophic turbulence}.  \jt{J.
  Fluid. Mech.}  \bvol{869},  \pg{R7}.

\bibitem[Kafiabad {\em et~al.\/}(2020)Kafiabad, Vanneste \&
  Young]{kafiabad_vanneste_young2020}
{\sc \au{Kafiabad, Hossein~A}, \au{Vanneste, Jacques} \& \au{Young, William~R}}
  \yr{2020}  \at{Wave-averaged geostrophic balance}.  \jt{arXiv preprint
  arXiv:2003.03389} .

\bibitem[Lelong \& Riley(1991)]{lelong}
{\sc \au{Lelong, M.-P} \& \au{Riley, J.~J.}} \yr{1991}  \at{Internal
  wave--vortical mode interactions in strongly stratified flows}.  \jt{J. Fluid
  Mech.}  \bvol{232},  \pg{1–19}.

\bibitem[Li \& Lindborg(2018)]{li-linb}
{\sc \au{Li, Q.} \& \au{Lindborg, E.}} \yr{2018}  \at{Weakly or strongly
  nonlinear mesoscale dynamics close to the tropopause?}  \jt{J. Atmos. Sci.}
  \bvol{75}~(4),  \pg{1215--1229}.

\bibitem[Lvov {\em et~al.\/}(2012)Lvov, Polzin \& Yokoyama]{lvov-et-al}
{\sc \au{Lvov, Y.~V.}, \au{Polzin, K.~L.} \& \au{Yokoyama, N.}} \yr{2012}
  \at{Resonant and near-resonant internal wave interactions}.  \jt{J. Phys.
  Oceanogr.}  \bvol{42}~(5),  \pg{669--691}.

\bibitem[{M{\"u}ller}(1976)]{muller76}
{\sc \au{{M{\"u}ller}, P.}} \yr{1976}  \at{On the diffusion of momentum and
  mass by internal gravity waves}.  \jt{J. Fluid Mech.}  \bvol{77}~(4),
  \pg{789--823}.

\bibitem[{M{\"u}ller}(1977)]{muller77}
{\sc \au{{M{\"u}ller}, P.}} \yr{1977}  \at{Spectral features of the energy
  transfer between internal waves and a larger-scale shear flow}.  \jt{Dynam.
  Atmos. Oceans}  \bvol{2}~(1),  \pg{49--72}.

\bibitem[Müller {\em et~al.\/}(1986)Müller, Holloway, Henyey \&
  Pomphrey]{mull-et-al}
{\sc \au{Müller, P.}, \au{Holloway, G.}, \au{Henyey, F.} \& \au{Pomphrey, N.}}
  \yr{1986}  \at{Nonlinear interactions among internal gravity waves}.
  \jt{Rev. Geophys.}  \bvol{24}~(3),  \pg{493--536}.

\bibitem[Nastrom \& Gage(1985)]{nast-gage}
{\sc \au{Nastrom, G.~D.} \& \au{Gage, K.~S.}} \yr{1985}  \at{A climatology of
  atmospheric wavenumber spectra of wind and temperature observed by commercial
  aircraft}.  \jt{J. Atmos. Sci.}  \bvol{42}~(9),  \pg{950--960}.

\bibitem[Nazarenko(2011)]{naza}
{\sc \au{Nazarenko, S.}} \yr{2011} {\em Wave Turbulence\/}, 1st edn.
  \publ{Springer}.

\bibitem[Olbers {\em et~al.\/}(2012)Olbers, Willebrand \& Eden]{olbers}
{\sc \au{Olbers, D.}, \au{Willebrand, J.} \& \au{Eden, C.}} \yr{2012} {\em
  Ocean Dynamics\/}, 1st edn.  \publ{Springer}.

\bibitem[Onuki(2020)]{onuki2020quasi}
{\sc \au{Onuki, Y.}} \yr{2020}  \at{Quasi-local method of wave decomposition in
  a slowly varying medium}.  \jt{J. Fluid Mech.}  \bvol{883}.

\bibitem[Powell \& Vanneste(2005)]{powe-v}
{\sc \au{Powell, J.} \& \au{Vanneste, J.}} \yr{2005}  \at{Transport equations
  for randomly perturbed {H}amiltonian systems, with application to {R}ossby
  waves}.  \jt{Wave Motion}  \bvol{42},  \pg{289--308}.

\bibitem[Qiu {\em et~al.\/}(2018)Qiu, Chen, Klein, Wang, Torres, Fu \&
  Menemenlis]{qiu-et-al-2018-seasonality}
{\sc \au{Qiu, B.}, \au{Chen, S.}, \au{Klein, P.}, \au{Wang, J.}, \au{Torres,
  H.}, \au{Fu, L.-L.} \& \au{Menemenlis, D.}} \yr{2018}  \at{Seasonality in
  transition scale from balanced to unbalanced motions in the world ocean}.
  \jt{J. Phys. Oceanogr.}  \bvol{48}~(3),  \pg{591--605}.

\bibitem[Rocha {\em et~al.\/}(2016)Rocha, Chereskin, Gille \&
  Menemenlis]{roch-et-al16}
{\sc \au{Rocha, C.~B.}, \au{Chereskin, T.~K.}, \au{Gille, S.~T.} \&
  \au{Menemenlis, D.}} \yr{2016}  \at{{Mesoscale to Submesoscale Wavenumber
  Spectra in Drake Passage}}.  \jt{J. Phys. Oceanogr.}  \bvol{46}~(2),
  \pg{601--620}.

\bibitem[Ryzhik {\em et~al.\/}(1996)Ryzhik, Papanicolaou \& Keller]{ryzhik}
{\sc \au{Ryzhik, L.}, \au{Papanicolaou, G.} \& \au{Keller, J.~B.}} \yr{1996}
  \at{Transport equations for elastic and other waves in random media}.
  \jt{Wave Motion}  \bvol{24}~(4),  \pg{327 -- 370}.

\bibitem[Savva(2020)]{savva2020}
{\sc \au{Savva, M. A.~C.}} \yr{2020}  \at{Inertia-gravity-waves in geostrophic
  turbulence}. PhD thesis, University of Edinburgh, Edinburgh.

\bibitem[Savva \& Vanneste(2018)]{savva_vanneste_2018}
{\sc \au{Savva, M. A.~C.} \& \au{Vanneste, J.}} \yr{2018}  \at{Scattering of
  internal tides by barotropic quasigeostrophic flows}.  \jt{J. Fluid. Mech.}
  \bvol{856},  \pg{504–530}.

\bibitem[Torres {\em et~al.\/}(2018)Torres, Klein, Menemenlis, Qiu, Su, Wang,
  Chen \& Fu]{torres2018partitioning}
{\sc \au{Torres, H.~S.}, \au{Klein, P.}, \au{Menemenlis, D.}, \au{Qiu, B.},
  \au{Su, Z.}, \au{Wang, J.}, \au{Chen, S.} \& \au{Fu, L.-L.}} \yr{2018}
  \at{Partitioning ocean motions into balanced motions and internal gravity
  waves: A modeling study in anticipation of future space missions}.  \jt{J.
  Geophys. Res. Oceans}  \bvol{123}~(11),  \pg{8084--8105}.

\bibitem[Vanneste(2013)]{Vanneste2013BalanceFlows}
{\sc \au{Vanneste, J.}} \yr{2013}  \at{Balance and spontaneous wave generation
  in geophysical flows}.  \jt{Annu. Rev. Fluid Mech.}  \bvol{45}~(1),
  \pg{147--172}.

\bibitem[Villani(2008)]{villani2008}
{\sc \au{Villani, C.}} \yr{2008}  \at{{H-theorem and beyond: Boltzmann's
  entropy in today's mathematics}}.  \jt{\emph{In} {Boltzmann’s Legacy},
  \emph{Gallavoti, G., Reiter, W.L., Yngvason, J., Eds., EMS Publishing House,
  Z\"urich, Switzerland}}  \pg{pp. pp. 129--143}.

\bibitem[Wagner {\em et~al.\/}(2017)Wagner, Ferrando \&
  Young]{wagner_ferrando_young_2017}
{\sc \au{Wagner, G.~L.}, \au{Ferrando, G.} \& \au{Young, W.~R.}} \yr{2017}
  \at{An asymptotic model for the propagation of oceanic internal tides through
  quasi-geostrophic flow}.  \jt{J. Fluid Mech.}  \bvol{828},  \pg{779–811}.

\bibitem[Wagner \& Young(2015)]{Wagner2015AvailableFlow}
{\sc \au{Wagner, G.~L.} \& \au{Young, W.~R.}} \yr{2015}  \at{{Available
  potential vorticity and wave-averaged quasi-geostrophic flow}}.  \jt{J.
  Fluid. Mech.}  \bvol{785},  \pg{401--424}.

\bibitem[Waite \& Bartello(2006)]{wait-bart06b}
{\sc \au{Waite, M.~L.} \& \au{Bartello, P.}} \yr{2006}  \at{The transition from
  geostrophic to stratified turbulence}.  \jt{J. Fluid. Mech.}  \bvol{568},
  \pg{89–108}.

\bibitem[Ward \& Dewar(2010)]{ward}
{\sc \au{Ward, M.~L.} \& \au{Dewar, W.~K.}} \yr{2010}  \at{Scattering of
  gravity waves by potential vorticity in a shallow-water fluid}.  \jt{J. Fluid
  Mech.}  \bvol{663},  \pg{478–506}.

\bibitem[Warn(1986)]{warn1986statistical}
{\sc \au{Warn, T.}} \yr{1986}  \at{Statistical mechanical equilibria of the
  shallow water equations}.  \jt{Tellus A}  \bvol{38}~(1),  \pg{1--11}.

\bibitem[Watson(1985)]{watson1984-internalWaves}
{\sc \au{Watson, K.~M.}} \yr{1985}  \at{Interaction between internal waves and
  mesoscale flow}.  \jt{J. Phys. Oceanogr.}  \bvol{15},  \pg{1296--1311}.

\end{thebibliography}

\end{document}